\documentclass[twocolumn]{aastex62}

\newcommand{\swiftxrt}{\textit{Swift/XRT}\xspace}
\newcommand{\ixpe}{\textit{IXPE}\xspace}

\newcommand{\maxi}{{MAXI}\xspace}
\newcommand{\nicer}{{NICER}\xspace }
\newcommand{\nustar}{\textit{NuSTAR}\xspace}

\newcommand{\degree}{$^\circ$\xspace}

\newcommand{\source}{GX~339--4\xspace}

\newcommand{\PDe}{\rm{P}} 
\newcommand{\PAe}{\theta} 
\newcommand{\PD}[1]{$\PDe_{\rm{#1}}$}
\newcommand{\PA}[1]{$\PAe_{\rm{#1}}$}
\newcommand{\PDX}{\PD{X}}
\newcommand{\PAX}{\PA{X}}
\newcommand{\PDXdisk}{\PD{X, disk}}  
\newcommand{\PDXnth}{\PD{X, nth}}

\newcommand{\PDXrelline}{\PD{X, relline}}
\newcommand{\PAXdisk}{\PA{X, disk}}  
\newcommand{\PAXnth}{\PA{X, nth}}

\def\heasoft{\textit{HEAsoft}\xspace}

\usepackage{graphicx}
\usepackage{amsmath}
\usepackage{tabularx}
\usepackage{booktabs} 
\usepackage{color}
\usepackage{soul}
\usepackage{cleveref}
\usepackage{xspace} 
\usepackage{xcolor}

\usepackage{multirow}
\usepackage{txfonts}
\usepackage{caption}
\usepackage{subcaption}

\usepackage{natbib}
\bibpunct{(}{)}{;}{a}{}{,}

\graphicspath{{./}{}}

\revised{August 9, 2024}
\submitjournal{ApJ}

\shorttitle{GX 339-4 polarization}
\shortauthors{Mastroserio et al.}

\begin{document}

\title{X-ray and optical polarization aligned with the radio jet ejecta in GX~339--4}

\correspondingauthor{Guglielmo Mastroserio}
\email{guglielmo.mastroserio@gmail.com}

\author[0000-0003-4216-7936]{G. Mastroserio}
\affil{Dipartimento di Fisica, Universit`a degli Studi di Milano, Via Celoria 16, I-20133 Milano, Italy}

\author[0000-0003-2743-6632]{B. De Marco}
\affil{Departament de Fis\'{i}ca, EEBE, Universitat Polit\`ecnica de Catalunya, Av. Eduard Maristany 16, 08019 Barcelona, Spain}

\author[0000-0003-1285-4057]{M. C. Baglio}
\affil{Istituto Nazionale di Astrofisica, Osservatorio Astronomico di Brera, via E.\,Bianchi 46, 23807 Merate (LC), Italy}

\author[0000-0002-0426-3276]{F. Carotenuto}
\affil{Astrophysics, Department of Physics, University of Oxford, Keble Road, Oxford OX1 3RH, UK}

\author[0000-0003-1533-0283]{S. Fabiani}
\affil{INAF-IAPS via del Fosso del Cavaliere, 100, 00133 Rome, Italy}

\author[0000-0002-7930-2276]{T.D. Russell}
\affil{INAF, Istituto di Astrofisica Spaziale e Fisica Cosmica, Via U. La Malfa 153, I-90146 Palermo, Italy}

\author{F. Capitanio}
\affil{INAF-IAPS via del Fosso del Cavaliere, 100, 00133 Rome, Italy}

\author[0000-0002-6447-3603]{Y. Cavecchi}
\affil{Departament de Fis\'{i}ca, EEBE, Universitat Polit\`ecnica de Catalunya, Av. Eduard Maristany 16, 08019 Barcelona, Spain}

\author{S. Motta}
\affil{Istituto Nazionale di Astrofisica, Osservatorio Astronomico di Brera, via E.\,Bianchi 46, 23807 Merate (LC), Italy}

\author[0000-0002-3500-631X]{D. M. Russell}
\affil{Center for Astrophysics and Space Science (CASS), New York University Abu Dhabi, PO Box 129188, Abu Dhabi, UAE}

\author[0000-0003-0079-1239]{M. Dov\v{c}iak}
\affil{Astronomical Institute of the Czech Academy of Sciences, Bo\v{c}n\'{i} II 1401, 14100 Praha 4, Czech Republic}

\author{M. Del Santo}
\affil{INAF, Istituto di Astrofisica Spaziale e Fisica Cosmica, Via U. La Malfa 153, I-90146 Palermo, Italy}

\author[0000-0003-0168-9906]{K. Alabarta}
\affil{Center for Astrophysics and Space Science (CASS), New York University Abu Dhabi, PO Box 129188, Abu Dhabi, UAE}

\author{A. Ambrifi}
\affil{ Instituto de Astrofísica de Canarias, E-38205 La Laguna, Tenerife, Spain}
\affil{Departamento de Astrofísica, Universidad de La Laguna, E-38206 La Laguna, Tenerife, Spain}

\author{S. Campana}
\affil{INAF, Istituto di Astrofisica Spaziale e Fisica Cosmica, Via U. La Malfa 153, I-90146 Palermo, Italy}

\author[0000-0002-0752-3301]{P. Casella}
\affil{INAF-Osservatorio Astronomico di Roma, Via Frascati 33, I-00076, Monte Porzio Catone (RM), Italy }

\author[0000-0001-9078-5507]{S. Covino}
\affil{INAF, Istituto di Astrofisica Spaziale e Fisica Cosmica, Via U. La Malfa 153, I-90146 Palermo, Italy}

\author[0000-0003-4795-7072]{G. Illiano}
\affil{INAF-Osservatorio Astronomico di Roma, Via Frascati 33, I-00076, Monte Porzio Catone (RM), Italy }
\affil{Tor Vergata University of Rome, Via della Ricerca Scientifica 1, I-00133 Roma, Italy}
\affil{Sapienza Università di Roma, Piazzale Aldo Moro 5, I-00185 Rome, Italy}

\author[0000-0003-0172-0854]{E.~Kara}
\affil{MIT Kavli Institute for Astrophysics and Space Research, Massachusetts Institute of Technology, Cambridge, MA 02139, USA}

\author[0000-0002-6421-2198]{E.~V. Lai}
\affil{INAF-Osservatorio Astronomico di Cagliari, via della Scienza 5, I-09047, Selargius (CA), Italy }

\author{G. Lodato}
\affil{Dipartimento di Fisica, Universit`a degli Studi di Milano, Via Celoria 16, I-20133 Milano, Italy}

\author{A. Manca}
\affil{Dipartimento di Fisica, Università degli Studi di Cagliari, SP Monserrato-Sestu, KM 0.7, Monserrato, 09042, Italy}

\author{I. Mariani}
\affil{INAF, Istituto di Astrofisica Spaziale e Fisica Cosmica, Via U. La Malfa 153, I-90146 Palermo, Italy}

\author[0000-0001-5674-4664]{A. Marino}
\affil{Institute of Space Sciences (ICE, CSIC), Campus UAB, Carrer de Can Magrans s/n, 08193, Barcelona, Spain}
\affil{Institut d'Estudis Espacials de Catalunya (IEEC), 08860 Castelldefels (Barcelona), Spain}

\author{C. Miceli}
\affil{Dipartimento di Scienze Fisiche ed Astronomiche, Università di Palermo, via Archirafi 36, 90123 Palermo, Italy}
\affil{INAF, Istituto di Astrofisica Spaziale e Fisica Cosmica, Via U. La Malfa 153, I-90146 Palermo, Italy}
\affil{IRAP, Universitè de Toulouse, CNRS, UPS, CNES, 9, avenue du Colonel Roche BP 44346 F-31028 Toulouse, Cedex 4, France}

\author{P. Saikia}
\affil{Center for Astrophysics and Space Science (CASS), New York University Abu Dhabi, PO Box 129188, Abu Dhabi, UAE}

\author[0000-0002-8808-520X]{A. W. Shaw}
\affil{Department of Physics \& Astronomy, Butler University, 4600 Sunset Ave, Indianapolis, IN 46208, USA}

\author[0000-0003-2931-0742]{J. Svoboda}
\affil{Astronomical Institute of the Czech Academy of Sciences, Bo\v{c}n\'{i} II 1401, 14100 Praha 4, Czech Republic}

\author[0000-0002-1481-1870]{F. M. Vincentelli}
\affil{Department of Physics and Astronomy, University of Southampton, SO17 1BJ, UK}

\author[0000-0002-1742-2125]{J.~Wang}
\affil{MIT Kavli Institute for Astrophysics and Space Research, Massachusetts Institute of Technology, Cambridge, MA 02139, USA}



\begin{abstract}
We present the first X-ray polarization measurements of \source. IXPE observed this source twice during its 2023-2024 outburst, once in the soft-intermediate state and again during a soft state. 
The observation taken during the intermediate state shows significant ($4\sigma$) polarization degree \PDX{} = $1.3\%\pm 0.3\%$ and polarization angle \PAX{} = -74\degree $\pm$ 7\degree only in the $3-8$~keV band. FORS2 at VLT observed the source simultaneously detecting optical polarization in the \textit{B}, \textit{V}, \textit{R}, \textit{I} bands (between $\sim0.1\%$ and $\sim0.7\%$), all roughly aligned with the X-ray polarization. 
We also  detect a discrete jet knot from radio observations taken later in time; this knot would have been ejected from the system around the same time as the hard-to-soft X-ray state transition and a bright radio flare occurred $\sim3$ months earlier. The proper motion of the jet knot provides a direct measurement of the jet orientation angle on the plane of the sky at the time of the ejection. We find that both the X-ray and optical polarization angles are aligned with the direction of the ballistic jet.

\end{abstract}



\section{Introduction}
Black hole X-ray binaries (BHXBs) are binary systems where a stellar-mass black hole accretes matter from a companion star. 
In the sub-class of these objects that hosts a low-mass companion star, material is transferred from the star to the black hole via Roche-lobe overflow, forming an accretion disk around the black hole. 
The quasi-totality of these systems is X-ray transient, i.e. they alternate months-long X-ray bright outbursts to longer quiescence periods when they are either not detected or very X-ray faint \citep{Tetarenko2016}. 
During the outburst phase, the X-ray radiation mainly consists of a multi-color disk-blackbody component from the geometrically thin, optically-thick accretion disk \citep{Shakura1973, Novikov1973} and a Comptonization component which is due to disk photons being up-scattered by hot electrons confined in a plasma region close to the black hole \citep[i.e., the corona,][]{Thorne1975}. 
A portion of the latter emission is reprocessed by the disk emitting a so-called reflection component \citep{Fabian1989}. 
During their outbursts, the majority of these systems follow a characteristic evolution through three main states defined by the X-ray luminosity and the hardness ratio of the energy spectrum \citep[e.g.][]{Belloni2010}. 
From the quiescence phase, they enter the \textit{hard state} increasing their X-ray luminosity with a relatively hard X-ray spectrum, dominated by the coronal emission.  
As the outburst progresses, the energy spectrum becomes softer \cite[over timescales of days;][]{Done2007} and the X-ray luminosity slightly increases, as the source enters an intermediate state. 
During this \textit{intermediate state}, both the disk and coronal emissions contribute significantly to the spectrum.  
However, based on the X-ray timing properties \citep[e.g.][]{Belloni2005}, this state can be further sub-classified in a hard-intermediate state (HIMS) and a soft-intermediate state (SIMS).  
The source continues to soften, transitioning towards a soft X-ray state.
Once the source is in the \textit{soft state}, it begins to slowly decay in X-ray brightness while the shape of its spectrum remains relatively constant, dominated by the disk component. 
Finally, the source transitions back to the hard state, tracing a hysteresis pattern in the hardness intensity diagram (HID) \citep[][]{Homan2001}. 

During their outbursts, BHXBs launch two types of jets, both observed in the radio band: a steady, compact jet is observed in the hard X-ray state, and a short-lived, transient jet arises during the hard-to-soft transition
\citep[e.g.,][]{Fender2004}. The transient jet is composed of discrete knots of synchrotron-emitting plasma that are launched out from the system. These ejecta produce bright radio flares as they propagate outwards along the jet axis (at the time of launch; \citealt{Miller-Jones2019}) at speeds that can approach the speed of light \citep[e.g.,][]{Mirabel1994,Tingay1995,Russell2019,Bright2020,Carotenuto2021}.

The timing properties of the X-ray emission change during the outburst. The source increases its root mean square (rms) X-ray variability
as the source brightens during the hard state, reaching a maximum at the beginning of the hard-to-soft transition \citep{MunosDarias2011}. 
The variability decreases slightly during the transition, dropping almost to zero during the soft state. 
Some of the most intriguing features are the X-ray quasi-periodic oscillations (QPOs) that appear in the hard and intermediate states. These QPOs have a characteristic timescale of $\sim 0.01-0.1$~Hz during the hard states, that move to a few Hz during the intermediate state \citep[][for a review]{Ingram2019b}. 
QPOs are divided into three classes: type-C QPOs,  which are detected in all states, type-B QPOs which are detected only in the soft-intermediate state and usually just before the transition to the soft state, 
and type-A QPOs, which are weaker and observed much more rarely \citep{Casella2005}. 

The two main components of the X-ray emission (disk and corona) are expected to be polarized with a linear polarization fraction (\PD{}) highly dependent on the viewing angle.
Photons emerging from a multi-upscattering inverse Compton process are expected to have a net polarization angle (\PA{}) parallel to the minor axis of the emitting region \citep{Poutanen1996, Schnittman2010}. 
The thermal emission from the thin accretion disk is expected to be polarized perpendicular to the plane of the disk \citep{Chandrasekhar1960, Sobolev1963, Dovciak2008, Schnittman2009}. 
The recently launched Imaging X-ray Polarimetry Explorer  \citep[\ixpe;][]{Weisskopf2022} has detected significant polarization in several BHXBs. 
Cyg~X-1 was the first black hole that showed a significant X-ray polarization measured by \ixpe. 
That system was observed during a hard state, where a 4$\%$ \PDX{} was measured, with a \PAX{} aligned with the jet axis \citep{Krawczynski2022}. 
Such an alignment constrains the geometry of the corona to be horizontally extended in the plane of the disk \citep[but see also][for alternative explanations]{Moscibrodzka2023, Dexter2024}. 
Recently \ixpe observed Swift~J1727.8$-$1613 during its intermediate state \citep{Veledina2023, Ingram2024}. During the hard-to-soft transition, the \PDX{} slightly decreased from $\sim 4\%$ to $\sim 3\%$ while the \PAX{} stayed constant. 
Importantly, the \PAX{} was aligned with the jet direction \citep{Wood2024arXiv}. 
\ixpe has also observed a few BHXBs in the soft state.
LMC~X--1 \citep{Podgorny2023}, 4U~1957$+$115 \citep{Marra2024}, LMC~X--3 \citep{Svoboda2024a}, 4U~1630$-$47 \citep{RodriguezCavero2023,Ratheesh2024} all exhibit significant \PDX{} ranging from $\sim1\%$ up to $\sim8\%$. 
While the sources with relatively low \PDX{} ($< 3 \%$) are easier to explain, reconciling the high \PDX{} of LMC~X-3 and 4U~1630$-$47 ($\sim4\%$ and $\sim8\%$ respectively) with theoretical models is more complicated considering the relatively low source inclination, requiring returning radiation from the accretion disk \citep{Schnittman2009, Taverna2021}.
Cyg~X--1 was also observed in a soft X-ray state and its \PDX{} decreased to $\sim2\%$ compared to the hard state while \PAX{} stayed constant \citep[][]{Steiner2024}. 

\subsection{\source\ 2023-2024 outburst}
\source\ is a low mass X-ray binary (LMXB) discovered in 1972 \citep{Markert1973}. 
The source goes into full-outburst\footnote{LMXBs, including \source, may also go through `failed' outbursts, where the source does not transit through all of the states, often remaining in a hard state only \citep{Tetarenko2016}.} every 2-3 years, although there does not appear to be any specific periodicity \citep{Alabarta2021}, moving through all of the X-ray states \citep[e.g.][]{Plant2014}. 
For these reasons \source\ is often considered the archetypal LMXB, often used to explain typical source behaviors and as a comparison source. 
Despite many outbursts, the distance, inclination, and mass of \source\ have not been precisely determined. 
\citet{Zdziarski2019} recently revised some of these parameters, claiming a narrower range for inclination $\sim$40\degree -- 60\degree \citep[compared to $\sim$37\degree -- 78\degree of][]{Heida2017} resulting in a distance to the source of $\sim8-12$~kpc, and a mass between $4-11 M_\odot$ (which is slightly higher than the previous constraint of $2.3-9.5 M_\odot$; \citealt{Heida2017}). 

From mid-August 2023, 
the Las Cumbres Observatory detected an increase in the source's optical flux \citep{ATel2023Alabarta_gx339}.
In October 2023, the Monitor of All-sky X-ray Image (MAXI) measured an X-ray flux of $0.033\pm0.007$ photons cm$^{-2}$ s$^{-1}$ in the $2-10$~keV energy band \citep{2023ATelNegoro_gx339}, while MeerKAT detected the initial brightening of the radio jet \citep{2024ATelNyamai_a}. 
On January 25, 2024, \source\ began a hard-to-soft transition \citep{2024ATelNegoro, Atel_Nyamai}, which was then followed by a fading phase in the optical \citep{2024ATelAlabarta}.
In Appendix~\ref{app:gx_outbursts} we compare the 2023-2024 outburst with the last four outbursts of \source. 
In this paper, we analyse two \ixpe observations performed during the 2023-2024 outburst in the soft-intermediate state and in the soft state. 
Thanks to simultaneous observation with other X-ray telescopes, we were able to constrain the spectral-timing-polarimetric properties of the source.
With the optical and radio coverage we measured the optical polarization and the orientation of the radio jet. 


\section{The multi-wavelength campaign}\label{sec:Obs}
\subsection{X-ray data}
\label{subsec:xray_obs}
All the X-ray observations have been reduced using the \heasoft\ release 6.33. 
\paragraph{\textbf{\ixpe}} \ixpe observed the source twice, once during the SIMS state and once during the soft state (see Table~\ref{tab:all_obs} in Appendix~\ref{app:gx_outbursts} for the details). Hereafter, we will refer to these observations as epoch 1 and epoch 2. 
For both observations, we extracted the source region using the \textit{SAOImageDS9}\footnote{www.sites.google.com/cfa.harvard.edu/saoimageds9} software.
We used a 60\arcsec\ circle centered at the peak of the source counts for each \ixpe detector.
We applied the NEFF-weighting analysis \citep{Baldini2022} extracting the energy spectra of the Stokes parameters I, Q, and U using \textsc{Xselect} from \heasoft.
Finally, we produced the instrument responses using the ftool \textit{ixpecalcarf} together with the response \textit{ixpe$\_$d2$\_$20170101$\_$alpha075\_02.rmf}. 
In both epoch 1 and epoch 2 we neglected the background contribution since the source is sufficiently bright \citep[see][for a complete explanation on the background subtraction in \ixpe]{DiMarco2023}.
The spectra were not rebinned during the fit, although they have been rebinned for visual purposes only in this paper. 


\paragraph{\textbf{\swiftxrt}} \swiftxrt \citep{Burrows2005} observed \source\ during the \ixpe observations multiple times (see Table~\ref{tab:all_obs} for details).
The \swiftxrt data were reprocessed with the task \texttt{xrtpipeline}, included in the \heasoft\ package, 
and the latest version of the CALDB files were employed. 
Events with grades 0 were selected to reduce the effect of energy redistribution at low energies that is
known to affect \textit{XRT} data for bright, heavily absorbed sources\footnote{See http://www.swift.ac.uk/analysis/xrt/digest\_cal.php for details.}.
Since the count rate in Window Timing mode was higher than 100 cts/s,  we applied the pile-up correction procedure by filtering the event files with an annulus region with an inner radius of 6 pixels (the outer radius was 20 pixels),
then the spectra were extracted running the task \texttt{xrtproducts}.
We found instrumental features in the spectra near the gold edge (2.2 keV) and the silicon edge (1.84 keV) which are present in high signal-to-noise WT spectra\footnote{http://heasarc.gsfc.nasa.gov/docs/heasarc/caldb/swift/docs/xrt/SWIFT-XRT-CALDB-09\_v19.pdf} and should be taken into account during the fitting procedure.

\paragraph{\textbf{\nicer}} The Neutron Star Interior Composition Explorer \citep{Gendreau2016} monitored \source\ during the 2023-2024 outburst. The observations considered in this paper started on January 31, 2024. The details of the observations are summarized in Appendix~\ref{app:gx_outbursts} (Table~\ref{tab:all_obs}).
The X-ray Timing Instrument (XTI) observations were reduced using the \nicer software \textsc{NICERDAS} distributed with \heasoft, the calibration files release \texttt{20240206}, and updated geomagnetic data. 
Calibration and screening of the data were performed using the \texttt{nicerl2} task, limiting undershoot rate to $\leq 200\ \rm{cts/s}$ and overshoot rate to $\leq1\ \rm{ct/s}$. 
Focal plane modules (FPMs) 14 and 34 were filtered out, due to occasional increased detector noise. We extracted light curves using the task \texttt{nicerl3-lc}. 
We note that all \nicer data of epoch 2 were taken during orbit day, thus the result is severely affected by the optical light leak which \nicer has been experiencing since May 22, 2023. We verified that the undershoot rate always exceeds $600\ \rm{cts/s}$ in this epoch. For this reason, we could not use these data for the analysis. For self-consistency, we decided to fit the energy spectra using only \swiftxrt for both epochs.

\paragraph{\textbf{\nustar}} The Nuclear Spectroscopic Telescope Array \citep{Harrison2013} observed the source during both epoch 1 and epoch 2.  
The same reduction procedure was applied to both observations. 
We used the \texttt{nupipeline}\footnote{Adding the specific \texttt{statusexpr="STATUS==b0000xxx00xxxx000".}} routine distributed from \heasoft\   to produce the event files and with the calibration files \texttt{20240325}.
The extraction region was a circle of 80\arcsec\ centered on the source for both epochs. 
We ran \texttt{nuproducts} to produce the energy spectra and \texttt{ftgrouppha} to rebin them with \texttt{grouptype=optmin} which uses the \citet{Kaastra2016} optimal rebinning, but it requires a minimum of counts in each bin (30 in our case).  

\subsection{Optical observations}\label{subsec:opt_obs}
\source\ was observed with the FOcal Reducer/low dispersion Spectrograph 2 (FORS2) mounted on the Very Large Telescope (VLT; Cerro Paranal, Chile) on February 15, 2024, for a total of 1617 s ($\sim 26$ minutes), simultaneously to the first \ixpe observations reported in this work (Sec. \ref{subsec:xray_obs}). 
A Wollaston prism was placed in the instrument's light path, dividing the incoming radiation into two orthogonally polarized light beams (ordinary and extraordinary). 
A Wollaston mask prevented these beams from overlapping on the CCD.
Additionally, a rotating half-wave plate (HWP) was installed, enabling images to be captured at four different angles ($\Phi_{i}$) relative to the telescope axis: $\Phi_i = 22.5^{\circ}(i-1), i = 1, 2, 3, 4$. 
This step is crucial for obtaining a polarization measurement, as the images taken at $>2$ different angles must be combined to determine the level of linear polarization, as described in Appendix~\ref{app:optical_pol_data}.  
Four sets of images were taken with this configuration, one for each available optical band ($I\_BESS$$+77$, $R\_SPECIAL$$+76$, $v\_HIGH$$+114$, $b\_HIGH$$+113$; hereafter, \textit{I}, \textit{R}, \textit{V}, and \textit{B}), with exposure times of 20s, 15s, 10s and 25s for each image in the four different filters, respectively.
Two more sets of observations were then acquired with the same configuration on different dates: March 2 and March 16, 2024. These two observations bracket - and are intended to be almost simultaneous with - the \ixpe epoch 2 (March 8-10, 2024).
All three observations were performed under photometric conditions, with seeing in the range of 0.3-0.4''.
All images were processed by subtracting an average bias frame and dividing by a normalized flat field (see additional details of the analysis in Appendix~\ref{app:optical_pol_data}).
We report the results of our optical polarization analysis in  Table~\ref{tab:polla_results}.

\begin{table*}[]
\caption{Results of the VLT/FORS2 ($BVRI$ filters) polarimetric campaign. All the polarization levels and angles are corrected for instrumental polarization. The interstellar polarization has also been subtracted, by means of a group of reference stars in the field. Upper limits are indicated at a $99.97\%$ confidence level, the rest of the errors at 1$\sigma$ confidence. }            
\label{tab:polla_results}      
\centering                       
\begin{tabular}{c |c |c |c |c |c| c |c |c}       
\hline               
 \multirow{2}{*}{Date} &\multicolumn{2}{c|}{$B$} & \multicolumn{2}{c|}{$V$} & \multicolumn{2}{c|}{$R$} & \multicolumn{2}{c}{$I$} \\   
 


 &\PD{} (\%) & \PA{} ($^{\circ}$)& \PD{} (\%) & \PA{} ($^{\circ}$) & \PD{} (\%) & \PA{} ($^{\circ}$) & \PD{}( \%) & \PA{}( $^{\circ}$)\\
\hline                       
2024 Feb 15&$0.65^{+0.14}_{-0.15} $ &$129\pm 6 $  & $0.35\pm 0.10 $ & $124\pm 8$ &$0.67^{+0.08}_{-0.09} $   & $128\pm 3$  &  $0.17\pm 0.05 $ & $103\pm 8$   \\
\hline
2024 Mar 02 & $ 2.1\pm 0.2 $ & $87\pm 3 $ &  $0.4\pm 0.1  $ &$180\pm 8 $ & $0.76^{+0.08}_{-0.09} $ & $119\pm 3 $ & $0.24\pm0.05 $ & $148\pm 5 $ \\
\hline

2024 Mar 16& $0.5\pm 0.1 $ & $96\pm 7$ & $0.26^{+0.07}_{-0.08} $ & $139^{+7}_{-8}$ & $0.17^{+0.05}_{-0.06} $ & $124^{+9}_{-10}$ & $<0.21\% $& $88^{+15}_{-15}$ \\
\hline

\end{tabular}
\end{table*}

\subsection{Radio observations}
\label{subsec:radio_observations}

We observed \source\ with the Australia Telescope Compact Array (ATCA) on May 1st 2024 between 16:30 UT and 23:13 UT (under program C3362; PI Carotenuto). ATCA was in its most extended 6A configuration.
Data were recorded simultaneously at central frequencies of 5.5\,GHz and 9.0\,GHz, with 2\,GHz of bandwidth at each frequency. 
We used PKS~B1934$-$638 for bandpass and flux density calibration, and B1646--50 for the complex gain calibration. 
Data were flagged, calibrated, and imaged using standard procedures within the Common Astronomy Software Applications for radio astronomy (\textsc{casa}, version 5.1.2; \citealt{CASAteam2022}).
When imaging, we used a Briggs robust parameter of 0 to balance sensitivity and resolution \citep{Briggs1995}.


\section{Timing analysis}
\label{sec:Spec}

\begin{figure} 
\centering
\includegraphics[width=\linewidth]{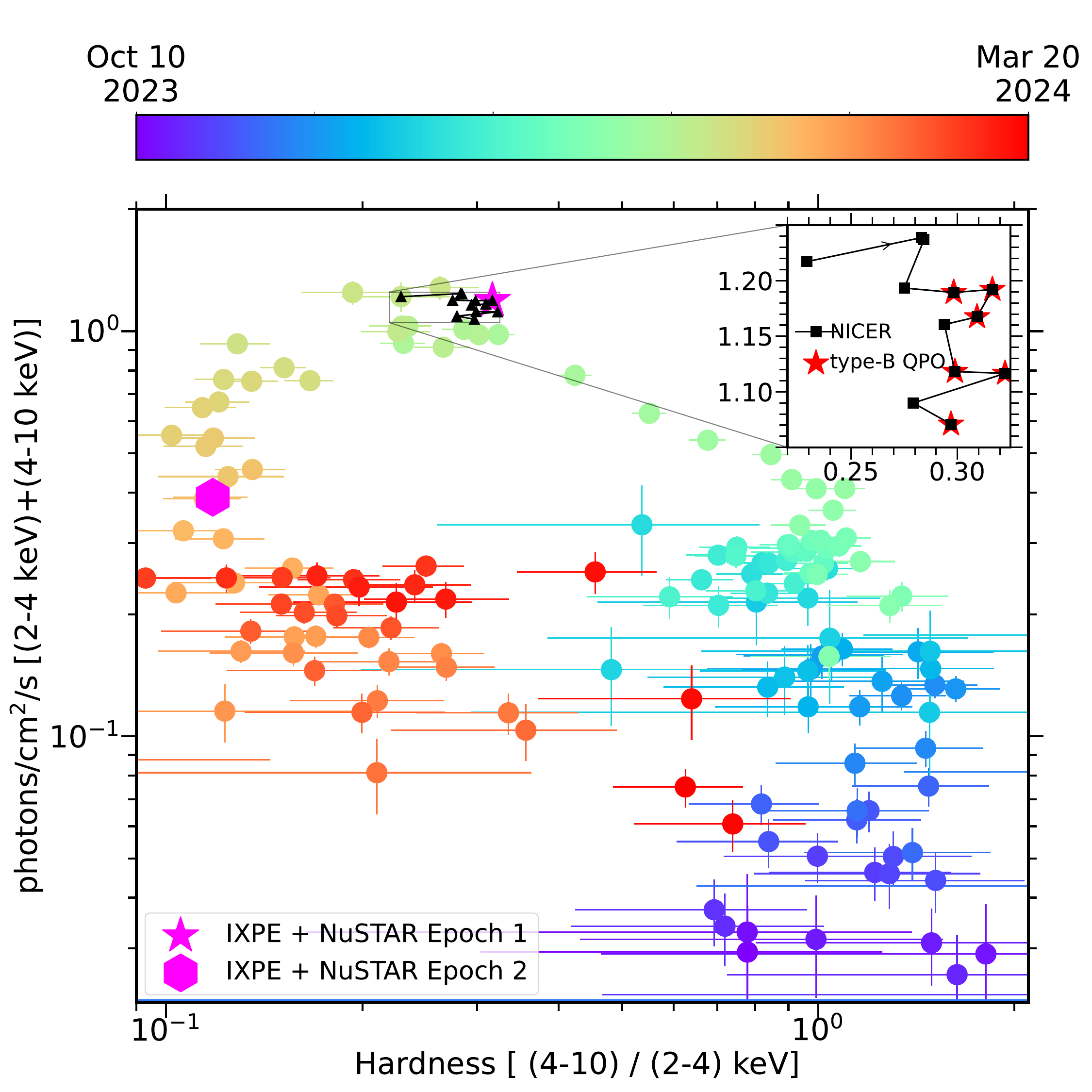}
\caption{Hardness intensity diagram (HID) of \source\ 2024 outburst computed by MAXI. The color scheme indicates the time order of the observations showing that the source starts in the hard state (purple points at the bottom right) and ends with the soft-to-hard transition (red points) following an anti-clockwise evolution. The black points are \nicer observations (from obsID~7702010107 to obsID~7702010118) with spectral hardness and count rate converted to the MAXI values (see the main text for details). The inset shows the zoom-in of the \nicer data and the red stars indicate the observations showing a type-B QPO. Finally, the 2 magenta symbols represent the first (star) and the second (hexagon) \ixpe observations simultaneous with \nustar and \swiftxrt.} \label{fig:HID}
\end{figure}

\subsection{HID and source evolution}
During the beginning of 2024, \source\ \nicer visibility was Sun-constrained, thus the observatory could not perform its typical daily observational campaign. 
However, \maxi\ and Swift/BAT \citep{Barthelmy2005} were able to observe the source almost every day, and we could follow the evolution of the spectral hardness during the outburst. 
Figure~\ref{fig:HID} shows the HID of the 2023-2024 outburst. 
The time evolution of the source is indicated by the different colors of the data points.
The right-top inset shows a zoom of \nicer observations (from obsID~7702010107 to obsID~7702010118). 
The two \ixpe observations of the source are indicated with the magenta symbols. 
The second \ixpe observation was performed on the same day as a \maxi\ pointing, thus we could easily place this observation on the HID. 
Unfortunately, the first \ixpe observation was performed during a gap in the \maxi\ monitoring. 
We used the \nicer data to track the evolution of the source during this period. 
The spectral hardness of \nicer data was computed using the same energy ranges used for \maxi. 
However, the two instruments have different effective areas, and both hardness and count rates are not trivial to compare. 
We decided to `calibrate' the \nicer count rates and hardness to match the \maxi\ values. 
We assigned to the \nicer obsID~7702010107 starting at MJD 60350.2924 (first point on the left in the inset of Figure~\ref{fig:HID}) the same count rate and hardness values as the \maxi\ observation at MJD 60350.5. 
We then applied a correction factor to \nicer data, to match the relative variations of hardness and rates observed in \maxi.

As it has often been observed in the past \citep[e.g.][]{Belloni2005}, \source\ did not smoothly transition to the soft state, but went through a number of back-and-forth transitions across the SIMS. We start to observe this behavior during the \maxi\ gap.
Past outbursts showed that, during this `hesitation' period, it is common to observe type-B QPOs, appearing and disappearing in the power density spectrum (PDS) \citep{Nespoli2003}.

\subsection{Fast X-ray variability and type-B QPO}
\label{sec:fast_var}
We analysed the fast X-ray variability of \source\ using \nicer and \nustar data, in order to better characterize the accretion state of the source and look for the presence of QPOs. 
For the \nicer data, we used $2-10$~keV light curves with a time bin of 0.0004~s and extracted a PDS for each observation using segments of 40~s length. 
For the \nustar data we used $3-80$~keV light curves with a time bin of 0.001~s and extracted the PDS using segments of 100~s length. 
We used the \texttt{stingray} software \citep{Huppenkothen2019,Bachetti2024_stingray} and the Fourier amplitude difference (FAD) to correct for deadtime \citep{Bachetti2018}. 
A small time bin allowed us to obtain a good sampling of the Poisson noise contribution to the PDS, which we fitted at frequencies $>300$ Hz with a constant model and subtracted out.

\begin{figure} 
\centering
\includegraphics[width=\linewidth]{ 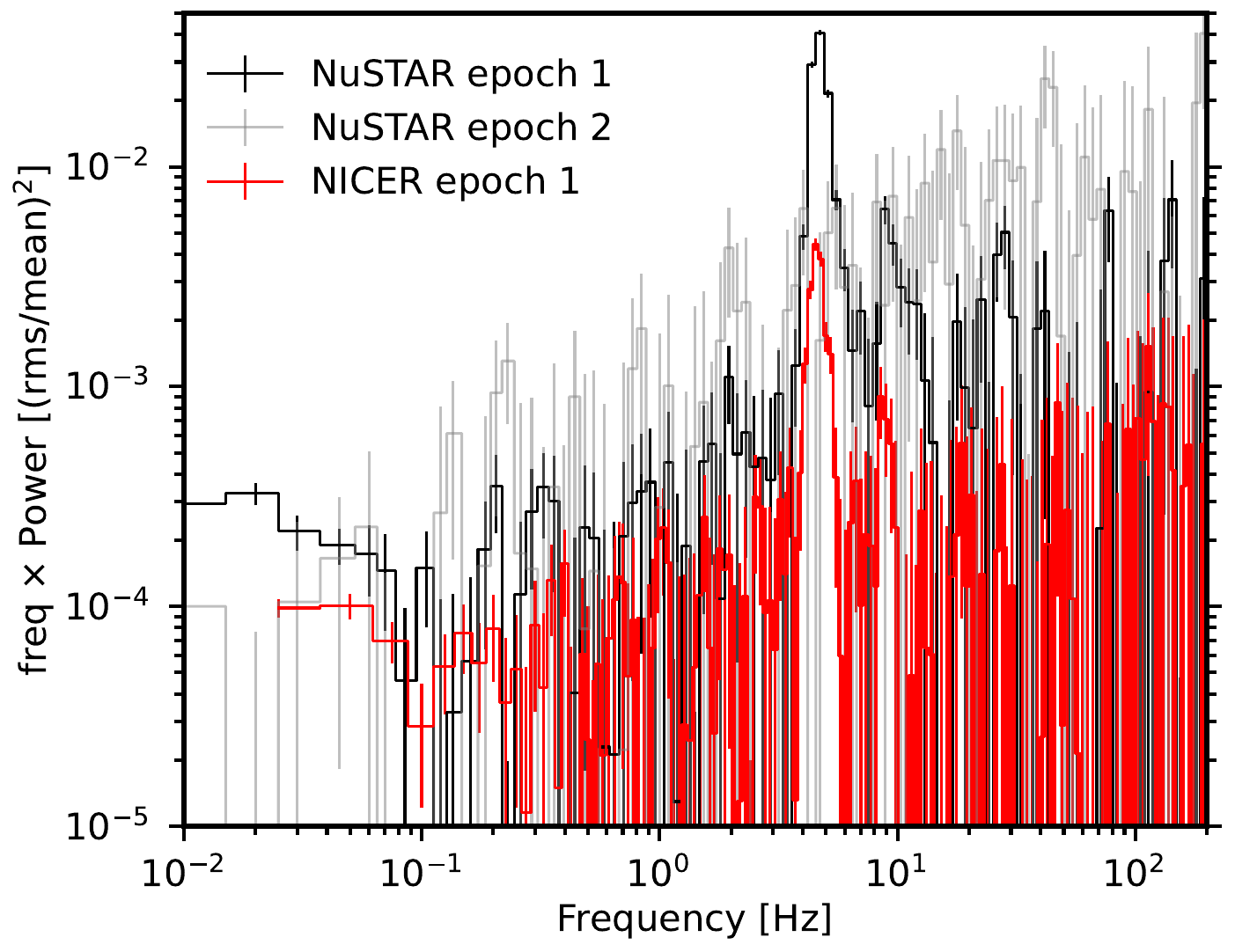}
\caption{The Poisson noise subtracted PDS obtained from \nustar (black and gray points for epoch 1 and 2, respectively) and \nicer (red points) data. The \nustar PDS is computed from the single observations during each epoch, while the \nicer PDS is computed by combining the 3 observations of epoch 1. The two PDSs have different powers due to the different energy ranges of the two instruments, the type-B QPO is stronger in \nustar due to the higher energy range considered.} 
\label{fig:PDS}
\end{figure}

A strong type-B QPO on top of a weak red noise continuum is detected in the \nustar and \nicer observations simultaneous to \ixpe epoch 1. 
In Figure~\ref{fig:PDS} we show the PDS obtained by combining the three \nicer observations simultaneous to epoch 1 (see Figure~\ref{fig:lc}).
Overplotted are the \nustar PDS simultaneous to epochs 1 and 2. 
More specifically, the QPO is observed in each of the three \nicer observations simultaneous to epoch 1, as well as in each of the \nustar orbits, suggesting that the source variability does not change significantly during epoch 1.
On the other hand, epoch 2 does not show significant fast variability during the \nustar observation (see gray PDS in Figure~\ref{fig:PDS}).
The intensive \nicer monitoring shows that the type-B QPO was not present over the 14 days preceding epoch 1, while it intermittently appeared and disappeared during the five days coverage following epoch 1. 
As previously pointed out, this behavior had been already seen in past outbursts \citep[][]{Nespoli2003}. 
\nicer observations containing a type-B QPO are marked with red stars in the inset of Figure~\ref{fig:HID}. 
Using a Lorentzian component to model the type-B QPO, we verified that its best-fit centroid frequency does not change significantly among \nicer observations. 
\nicer
The best-fit centroid frequency of the type-B QPO is $4.56\pm0.03$~Hz for \nicer combined data and $4.65\pm0.02$~Hz for \nustar data.

Given these results, and following the classification proposed in \cite{Belloni2005}, we infer that \source\ was likely in a soft-intermediate state during the first \ixpe observation. 
The lack of significant variability power in epoch 2, combined with the lower values of spectral hardness (Figure~\ref{fig:HID}), suggests that the source was in a soft state during the second \ixpe observation, as also confirmed by the results of our spectral analysis (Section~\ref{sec:spec_fit}).

\section{Spectro-Polarimetric Analysis}
\label{sec:spec_pol_fit}
We fitted the two epochs simultaneously using \ixpe, \nustar, and \swiftxrt data\footnote{We used \swiftxrt instead of \nicer in both epochs for consistency reasons (\nicer data during epoch 2 cannot be used due to instrumental issues, see  Section~\ref{subsec:xray_obs}).}. 
The latter includes three \swiftxrt observations in epoch 1, and the only \swiftxrt observation (obsID~00016552003) simultaneous to \nustar in epoch 2 (Table~\ref{tab:all_obs}). 
We first performed broad band fits of the energy spectrum (Section~\ref{sec:spec_fit}) using the \swiftxrt and \nustar data of the two epochs simultaneously. 
After constraining the best-fit spectral model, we added the \ixpe data (Section~\ref{sec:pol_fit}) in order to constrain the \PDX{} and \PAX{} of the source.
The errors on the parameter values in this Section are expressed at $90\%$ confidence if not specified otherwise.  

\subsection{Spectral fit}\label{sec:spec_fit}
Preliminary fits with a simple model (i.e. a multicolor disk-blackbody, \texttt{diskbb}, plus a Comptonization model, \texttt{nthcomp}) show that in epoch 1 both the disk and the Comptonization components contribute to the fit, while in epoch 2 the disk dominates the broadband emission, even though a dimmer Comptonization component is still required at high energies. 
Both epochs also show strong Fe~K$\alpha$ line residuals, even though there is no evidence of residuals in the Compton hump energy range. 
We note that this may happen because in these observations the source had a softer spectrum compared to the canonical hard state, and the illuminating spectrum reprocessed by the disk may be produced by the disk itself through returning radiation \citep{Connors2020, Connors2021}, thus lacking the hard energy photons that produce the typical Compton hump feature. 

Therefore, we decided to fit the data with the model \texttt{TBnew\_feo*(thcomp*kerrbb + relline)*smedge}. 
The choice of not including a more sophisticated reflection model \citep[such as the \texttt{relxill} model;][]{Garcia2014,Dauser2014} is dictated by the fact that such a model would overfit the Compton hump. 
Furthermore, the adopted model allows us to easily separate the contribution of the iron line component from the polarization data (Section \ref{sec:pol_fit}). 
Since the distance, inclination, and BH mass of \source\ are not well known, and we fix the distance to 10~kpc \citep{Zdziarski2019} and the mass to 10~$M_{\odot}$, we allow the inclination to vary between $\sim$30\degree -- 80\degree \citep{Heida2017}. 
Additional details on the spectral fits are given in Appendix~\ref{app:spectral_fit}.

According to our best-fit (reduced $\chi^2$ = 2461 / 2254), the inclination is pegged at the lower limit $i=30^{\circ}$. 
This value is slightly lower than the range of $\sim$40\degree -- 60\degree reported by \cite{Zdziarski2019}. 
The spin and the mass accretion rate are also free parameters of the fit, with the latter left free to vary between the two epochs.
We note that the spin value we find ($a=0.85\pm0.01$) is at risk of high degeneracy with other parameters of \texttt{kerrbb}. 
Moreover, the uncertainty we quote corresponds solely to the statistical errors and does not include any uncertainty due to model-dependent systematics. 
A robust estimate of the BH spin requires thorough testing of different models, possibly combining data from more than two observing epochs. 
However, this is beyond the scope of this work.
Here we are primarily interested in estimating the contribution of the main spectral components to the X-ray spectrum, to correctly interpret the polarization results (Section~\ref{sec:pol_fit}).  
The overall flux decreased between the two epochs, e.g. the $E=2-8$~keV flux is $F_{2-8\ keV}\sim 8.9 \times 10^{-9}$~ergs~cm$^{-2}$~s$^{-1}$ in epoch 1 and $F_{2-8\ keV}\sim 2.8 \times 10^{-9}$ ~ergs~cm$^{-2}$~s$^{-1}$ in epoch 2\footnote{The flux was estimated with the command \texttt{flux} in \texttt{xspec}.}. 
According to the results of our fit, this is driven by a significant reduction of the disk mass accretion rate, from $\dot{M}=1.06^{+0.13}_{-0.04}\times 10^{18} g/s$ in epoch 1 to $\dot{M}=0.45^{+0.02}_{-0.05}\times 10^{18} g/s$ in epoch 2. 
We also note that the covering fraction of the Comptonization model dropped from $0.17^{+0.02}_{-0.01}$ in epoch 1 to $0.029^{+0.002}_{-0.014}$ in epoch 2. 
This result is not surprising since the corona is expected to reduce its contribution as a source transitions to the soft state. 
The results of our spectral fits indicate that the spectrum is compatible with what is expected for a soft-intermediate state and a soft state during the first and second epochs, respectively. 
These results also agree with our conclusions from the analysis of the fast X-ray variability (Section~\ref{sec:fast_var}).

\subsection{Polarization fit}\label{sec:pol_fit}

\begin{figure} 
\centering
\includegraphics[width=\linewidth]{ 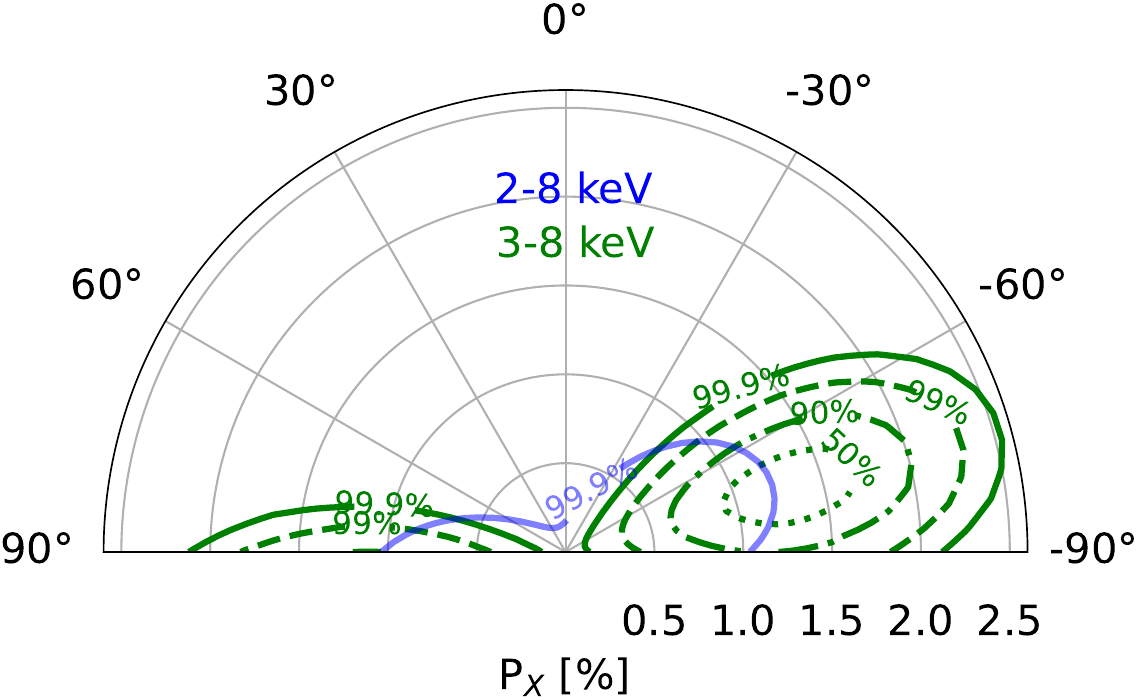}\vspace{3pt}
\includegraphics[width=\linewidth]{ 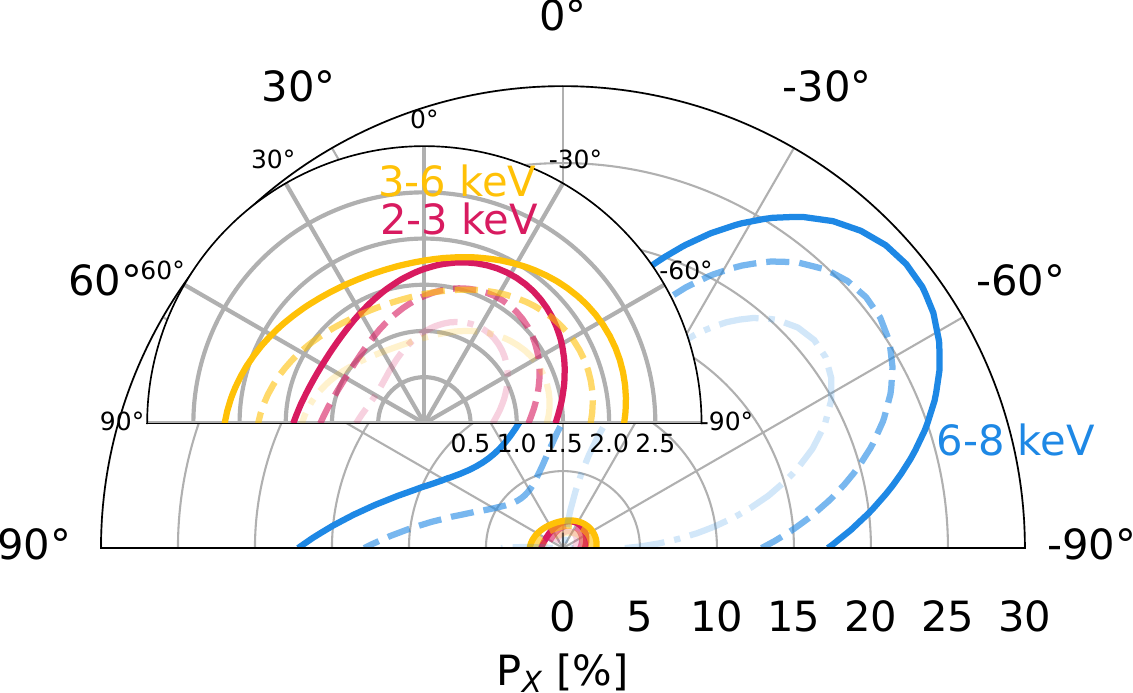}

\caption{The epoch 1 (top panel) and the epoch 2 (bottom panel) \PDX{} and \PAX{} contour levels from the simultaneous \ixpe, \swiftxrt and \nustar best fit (see Section~\ref{sec:spec_pol_fit} for details about the spectral-polarimetric fit). The \PDX{} and \PAX{} calculation of epoch 1 in the \ixpe $3-8$~keV energy band shows significant polarization (the 99.9\% contour green solid line closes before zero), while the full \ixpe band (blue solid line 99.9\% contour) does not show any significant polarization as in all the contour in epoch 2. The inset of the bottom panel is the zoom-in of the main plot. In the bottom panel the line styles are the same as indicated for the top panel.} 
\label{fig:polarization}
\end{figure}

We detected significant polarization in the $3-8$~keV energy range of epoch 1, while no significant polarization (above 3$\sigma$) was detected in epoch 2.

We included the \ixpe spectra to the fit, considering \ixpe epoch 1 first. 
The parameters of the joint epoch model were fixed to their best-fit values (see Table~\ref{tab:bestfit_par}), save from the calibration constants accounting for the different normalizations of different instruments and detectors.  
The spectra of the Stokes parameter \textit{I} require adding a gain shift model which allows the fit to shift the energies on which the response matrix is defined. 
We report the gain values in Table~\ref{tab:bestfit_par}. 
We verified that leaving the best-fit parameters free to vary after the addition of the \ixpe spectra does not cause the parameters to change significantly. 

We then fit the Stokes parameters \textit{Q} and \textit{U} of all the three detectors considering either the full \ixpe energy range ($E=2-8$~keV) or separating the \ixpe band in the two sub-ranges $E=2-3$~keV and $E=3-8$~keV. 
This choice is justified by the presence of the two main components in the spectrum, the disk and the Comptonization component, with the former dominating the softer energy range and supposed to be less polarised than the latter. 
The \texttt{polconst} model was applied to our best-fit model (found before considering \ixpe), thus differences in polarization among spectral components were not accounted for at this stage (see Appendix~\ref{app:spectral_fit} for additional details on the spectral-polarization analysis). 
Figure~\ref{fig:polarization} summarises our polarization results for epoch 1. 
We note that the 99.9\% contour level of the \ixpe full energy range does not close before the zero of \PDX, however, in the $3-8$~keV energy range we detect a significant polarization of $1.3\% \pm0.3\%$ (1$\sigma$ error). 
We calculated the significance with the \texttt{steppar} command in \texttt{xspec}, which shows that zero polarization is rejected with $\Delta\chi^2$ of 16.574, i.e. just above the 4$\sigma$ significance. 
The \PAX{} is also constrained to $-74$\degree$\pm7$\degree (1$\sigma$ error).

We followed the same procedure to perform the spectral-polarization fit of the second epoch.  
We could not detect any significant polarization either in the full \ixpe energy range or in any tested sub-energy range, only an upper limit at \PDX{}$<1.2\%$ (at 3$\sigma$ confidence level). 
We note that the most significant \PDX{} is measured in the $6-8$~keV \ixpe energy range (see bottom panel of Figure~\ref{fig:polarization}).
Only the 90\% confidence level contour closes before \PDX{} = 0, which can not be considered a detection. 
However, it is interesting to note that the \PAX{} of this contour is aligned with the \PAX{} of epoch 1. 

\section{Detection of a discrete jet ejection} \label{sec:radio_results}

\begin{figure} 
\centering
\includegraphics[width=\linewidth]{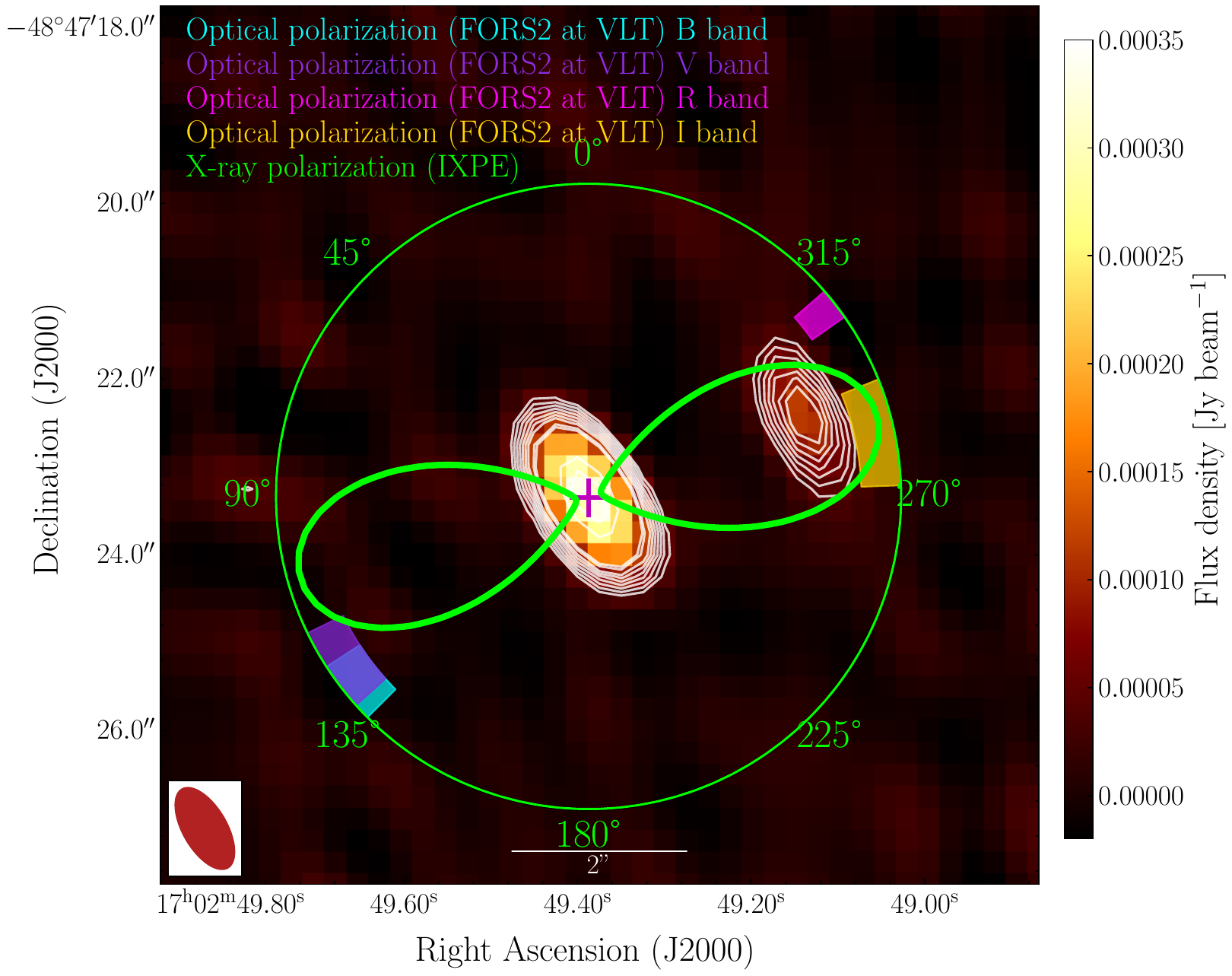}
\caption{ATCA 9 GHz image of \source\ showing the two radio sources, with the central source being associated with the core position of \source\ and the NE component being the discrete knot that was ejected from the system along the jet axis at a position angle of $-69.54$\degree $\pm$ $1.1$\degree East of North. The ATCA beam, shown in red on the bottom left corner, is $1.5\arcsec \times 0.7\arcsec$. The white radio contours start at three times the image noise rms, which is 10 $\mu$Jy/beam. \source\ position is marked with a magenta cross. We also display in green the \ixpe polarization 99.9\% contours in the $3-8$~keV which is the same solid line of Figure~\ref{fig:polarization} (\PDX{} =  $1.3\% \pm0.3\%$, 1$\sigma$ error).
The colour bands on the green circle indicate the optical polarization angle \PA{} of the four bands that we measure in epoch 1 (the values of the optical \PD{} are reported in Table~\ref{tab:polla_results}).} \label{fig:atca}
\end{figure}

In Figure~\ref{fig:atca} we present the ATCA image at 9~GHz (having a higher resolution than the 5.5~GHz map). 
The image shows two components: a stronger component located at the known position of \source, and a second component located to the NW of the first, at coordinates (J2000): R.A. = 17$^{\rm h}$02$^{\rm m}$49.139$^{\rm s}$ $\pm$ $0.02$\arcsec\ and Dec. = $-48$\degree$47$\arcmin$23.42$\arcsec $\pm$ 0.04\arcsec. The angular separation between the two components is $2.6\arcsec \pm 0.1\arcsec$.

The first component has a flux density of $400 \pm 6$ $\mu$Jy at 5.5 GHz and $356 \pm 8$ $\mu$Jy at 9 GHz, implying a radio spectral index $\alpha = -0.23 \pm 0.05$ (where the radio flux density follows $S_{\nu} \propto \nu^{\alpha}$), which is typical of the jets usually observed during the low hard state of this source \citep[e.g.][]{Corbel2013}.
The second component located $2.6\arcsec$ away has a flux density of $160 \pm 8$ $\mu$Jy at 5.5 GHz and of $122 \pm 6$ $\mu$Jy at 9 GHz, indicating $\alpha = -0.6 \pm 0.1$, typical of optically-thin ejecta launched from a BHXB around the hard-to-soft state transition \citep[e.g.][]{Corbel2002_xte}.

Based on their properties, we associate the first component with the core  of the jet, which had re-activated as the source returned to the hard state during the outburst decay \citep{Atel_Russell}. On the other hand, the second component is the result of a discrete ejection that occurred earlier during the outburst, most likely being ejected around the hard-to-soft state transition and having propagated outwards. This ejected knot remained radio bright as it interacted with its surroundings. 
Using the position of the ejected component and the core emission, we determine the jet orientation angle on the plane of the sky to be $-69.5$\degree $\pm$ $1.1$\degree East of North (at the time of the ejection). This jet orientation is fully consistent with the position angle of the ejecta reported in \cite{Gallo2004}. 
Given the large distance of \source \citep{Zdziarski2019}, resolving the compact jets to determine their position angle is particularly challenging \citep[e.g.,][]{2024arXiv240512370W}, as such, spatially resolving ejected jet components is crucial as it is often the only way to infer the jet orientation angle.
If the ejected component was launched at, or around, the start of a bright radio flare that was observed on January 31 \citep[MJD 60340,][]{Atel_Nyamai} the observed angular separation between the core and the ejection would imply a time-averaged proper motion of $\sim$28.5 mas d$^{-1}$, which is consistent with previous findings about this source \cite{Gallo2004}.

\section{Discussion}\label{sec:discussion}
We present the analysis of two \ixpe observations of the black-hole X-ray binary \source\ during its 2023-2024 outburst. The first and second \ixpe epochs were taken during the intermediate state and the soft state, respectively. Simultaneous \swiftxrt, \nicer and \nustar X-ray observations supported \ixpe during both epochs, while ATCA radio observations and VLT optical observations provided information on the multi-wavelength behavior of the source.

\subsection{Corona on the disk plane}\label{sec:sub_discussion_horiz_corona}

We have measured a $1.3\% \pm0.3\%$ (1$\sigma$ error) polarization fraction of \source\ in the $3-8$~keV energy band when the source was crossing the intermediate state. 
At the same time, a type-B QPO was detected in the light curve. 
Such a feature has been proposed to be linked with the launch of relativistic jets \citep[][]{Soleri2008, Fender2009, Miller-Jones2012, Russell2019, Russell2020b, Homan2020, Carotenuto2021, Wood2021}, although the connection is not completely clear and no physical model has been proposed yet to interpret it. 
Based on our results, it is reasonable to associate the jet ejection observed in the ATCA data with the radio flare detected by MeerKAT two weeks before \ixpe epoch 1. 

The measured direction of the jet on the plane of the sky is compatible with both the X-ray and optical polarization angles measured by \ixpe and FORS2 in epoch 1, which would position the corona perpendicular to the emitted jet.  
In order to confirm this interpretation, we need to prove that the measured \PDX{} is due to the coronal component. 
We fit the energy spectrum of epoch 1 with a \texttt{TBnew\_feo(diskbb + nthcomp + relline)*smedge} model, therefore we can assign the polarization model (\texttt{polconst}) to each component\footnote{We use the model \texttt{TBnew\_feo(polconst*diskbb + polconst*nthcomp + polconst*relline)*smedge} where each of the \texttt{polconst} has its own \PDX{} and \PAX{} parameter. }. 
Computing the 2D contours between the disk polarization (\PDXdisk) and the coronal polarization (\PDXnth)\footnote{\PDXrelline{} is fixed to zero.}, we find that the polarization could be caused by each of the two components independently of the assumed offset between their \PAX{}s (i.e. either aligned or perpendicular,
see Figures~\ref{fig:Xpol_components} and  \ref{fig:Xpol_components2}
in Appendix~\ref{app:spectral_fit}).
However, since the second \ixpe epoch, which corresponds to a disk-dominated spectrum, does not show any significant polarization, we speculate that the detected polarization in epoch 1 is likely to be ascribed to the Comptonization emission. 
If this is the case, the corona should be horizontally extended on the plane of the accretion disk, since photons emitted after multiple scatterings are polarized perpendicular to the accretion disk and the corona itself. 

In this scenario, the alignment of the optical polarization with the X-ray polarization suggests that the disk plane lies on the plane defined by the binary orbit. 
The optical polarization would then arise from Thomson scattering in the outer accretion disk, similarly to what has been reported for A0620--00, GRO~J1655--40 and Cyg~X--1 \citep{Dolan1989,Gliozzi1998,Krawczynski2022}.
We note that the optical polarization is very low ($<1\%$), as typically observed for BHXBs at relatively low inclinations as \source\ \citep[][]{Poutanen2018, Kravtsov2022}. 
This interpretation is also supported by the decreasing trend of the optical linear polarization towards the \textit{I}-band (i.e. the lowest optical frequency in our dataset) in all three VLT epochs (Table~\ref{tab:polla_results}).
 This behavior is in fact expected since the disk is typically more dominant at the highest optical frequencies \citep{Brown1978, dolan1984}.

\subsection{Corona geometry in the soft-intermediate state}\label{sec:sub_discussion_reverb}
Our X-ray polarization analysis, combined with results from optical and radio observations, suggests the corona to be horizontally extended in the soft-intermediate state. 
Observations of soft X-ray reverberation lags in \source\ during previous outbursts revealed the presence of short lags of a few milliseconds during the hard state \citep{DeMarco2015, DeMarco2017, Wang2020, Wang2022}. 
These lags have been shown to suddenly increase by a factor of $\sim10$ when the source moves to the hard-intermediate state, before its full transition to the soft state \citep{Wang2022}. 
If the increase of X-ray reverberation lag amplitude before the transition is entirely due to light travel time delays, the distance between the hard X-ray dissipation region and the disk must expand. 
A suggested solution is that the corona vertically extends or is partly ejected, possibly provoking shocks at large distances \citep{DeMarco2021,Wang2021,Wang2022}.  


Nonetheless, there are a few caveats that need to be considered. 
A long X-ray reverberation lag has not yet been observed in \source\ during the soft-intermediate state \citep[][a long lag in this state has been observed only in the very bright sources MAXI~J1820+070 and MAXI~J1348--630]{Wang2022}. 
We searched for soft X-ray lags in the analysed observations, but we could not detect any due to the very low variability of the source (the measured fractional rms was only a few percent). 
Moreover, due to visibility limitations, our campaign missed the hard-intermediate state of the 2023-2024 outburst, so we cannot directly compare our data to the results on X-ray reverberation reported by \citet{Wang2022} for the previous outburst of \source. 
Therefore, if we assume that the measured polarization is intrinsic to the corona, whatever the geometry of the corona at the beginning of the transition (hard-intermediate state), it should acquire a horizontal structure when reaching the soft-intermediate state, as inferred from X-ray polarization (but see discussion in Section~\ref{sec:sub_discussion_jet_pol}).
On the other hand, the requirement for a horizontally extended corona during the soft-intermediate state could be relaxed if the X-ray polarization originates from the reprocessing of the disk. 

We note that \citet{Peirano2023} performed a spectral-timing analysis of \source\ during the soft-intermediate state in the 2021 outburst, and they invoke a two-coronae structure in which a vertically extended corona explains the type-B QPO variability and lags, and a horizontally extended corona impacts the spectral shape.
This scenario \citep[proposed also for other sources in the soft-intermediate state;][]{Garcia2021, Ma2023} could reconcile the need for both a vertically structured corona to explain the X-ray reverberation lags and a horizontally extended corona required by the polarization. 

It is worth noting that from their X-ray polarimetric analysis of Swift J1727.8-1613 combined with the simultaneous detection of a long soft X-ray lag in the hard-intermediate state, \citet{Ingram2024} propose delays other than light-crossing time to significantly contribute to the observed lags \citep[e.g.][]{Salvesen2022}. 
In this context, several theoretical solutions have been proposed in recent years.
For example, \citet{Uttley2023arXiv} showed that, when properly accounting for mass accretion rate fluctuations starting in the accretion disk and propagating into the corona \citep[e.g.][]{Lyubarskii1997, Kotov2001}, the observed soft X-ray lags can be reproduced without the need for extremely large (neither radially nor vertically) coronae. 
On the other hand, \citet{Veledina2018} predicts soft lags from the interplay between two Comptonization components in a horizontally stratified corona which would cause a pivoting powerlaw \citep[see also][for a similar mechanism]{Mastroserio2018, Mastroserio2021}. 
Future monitoring allowing for simultaneous X-ray timing and polarimetric analysis throughout different accretion states will allow us to better test these hypotheses.

\subsection{Polarization from the jet}\label{sec:sub_discussion_jet_pol}

Alternative scenarios have been proposed to explain the alignment of \PAX{} with the direction of the jet. 
\citet{Dexter2024} proposed a solution in which the polarization is produced by scattering concentrated along the walls of a hollow-cone jet structure and the observer's line of sight falls inside the jet hollow-cone.
Since the jet cone angle is not expected to be very large observing a source at the right orientation is relatively unlikely. 
So far, all the sources that show significant polarization together with the detection of a ballistic jet have shown a  \PAX{} aligned with the jet direction. 
Moreover, the \citet{Dexter2024} model predicts quite a high polarization $\sim 4\%$, while we only detect $\sim1\%$. 

An alternative explanation is that the observed \PDX{} is produced by optically thin synchrotron emission from accelerated particles in the jet \citep[e.g.][]{Russell2014}. 
Following the calculation in \citet{Krawczynski2022}, the estimated \PDX{} from synchrotron emission is $< 8\%$ for a $\sim$30\degree inclination source \citep[see][for detailed calculations]{Lyutikov2005}. 
For Cyg~X--1, which might have a similar inclination angle to \source, the jet contribution to the soft X-ray flux is assumed to be $\lesssim 5\%$, resulting in an estimate of \PDX{} $< 0.4\%$ from the optically thin synchrotron jet. 
In the hard state of \source, the jet contribution could be $\sim$ten percent or less, 
and be produced by either the optically thin synchrotron from accelerated particles \citep[e.g.][]{Nowak2005,Gandhi2011,Russell2023}, or Compton scattering and/or synchrotron self-Compton emission from the base of the jet \citep[e.g.][]{Markoff2005,Connors2019b}. 
However, the optically thin synchrotron jet, seen clearly at infrared wavelengths in \source in the hard state, fades near the start of the state transition \citep[around the time of the transition from the hard state to the hard-intermediate state; e.g.][]{Homan2005b,Coriat2009,CadolleBel2011}, well before the radio emission from the jet is quenched. 
The X-ray contribution of the optically thin jet is therefore likely to be low during the soft-intermediate state. 

Similarly, the optical \PD{} measured during the first VLT epoch (i.e., in the soft-intermediate state; Table~\ref{tab:polla_results}) quasi-simultaneously with \ixpe is unlikely to have originated from optically thin synchrotron emission from the compact jet since this component is likely to have faded before this point along the transition. 
The systematic lower polarization measured in all bands in the third epoch of observations (i.e., during the soft state, when the jet is even more likely to be quenched; e.g., \citealt{Russell2020}) with respect to the first epoch (Table~\ref{tab:polla_results}) could suggest a weak but significant jet contribution in the first epoch. 
It has been shown that the jet can produce polarized light up to a few percent in the optical, depending on the level of ordering of the magnetic field lines at the base of the jet \citep[see e.g.][]{Russell2016, Shahbaz2019, Baglio2020}. 
\citet{Russell2011} demonstrated that the compact jet of \source\ produced a near-infrared (NIR) \PD{} of a few percent (a maximum of $2-3\%$) while in the hard state, with a \PA{} of $17^\circ\pm 1.6^{\circ}$, which is perpendicular to the radio jet axis and to the \ixpe and VLT \PA{} presented in this work. 
This result indicated that the magnetic field near the base of the jet was largely tangled in the hard state, with an average direction parallel to the jet axis.

If the optical \PD{} in the first epoch was due to the jet, considering the measured polarization angle, this could instead suggest that at the time of our observations the magnetic field at the base of the jet was perpendicular to its axis, similar to what was observed, e.g., in the case of a polarization flare from the BHXB V404~Cyg during its 2015 outburst \citep{shahbaz2016}. 
However, a more likely scenario could be optical polarization from electron scattering in the disk atmosphere \citep{Dolan1989,Gliozzi1998,Baglio2016, Kravtsov2022}. 
A more detailed analysis of the optical emission is needed to disentangle the different scenarios for the origin of the optical linear polarization and will be presented in a dedicated follow-up work (Baglio et al., in prep.).

\section{Summary and Conclusions}
We have analyzed a multi-wavelength campaign of \source\ during the 2023-2024 outburst performed with the \swiftxrt, \nicer, \nustar, and \ixpe X-ray telescopes, FORS2/VLT optical spectrograph and ATCA radio telescope. 
Here we summarize the results of our spectro-polarimetric timing analysis:
\begin{itemize}
    \item During the first \ixpe epoch we measured relatively soft energy spectra, which still showed some contribution from a hard non-thermal component, while the power density spectra displayed a clear type-B QPO at $4.56\pm0.03$ Hz for \nicer data and $4.65\pm0.02$ Hz for the \nustar data. 
    The QPO indicates that the source was in the soft-intermediate state during epoch 1. 
    During the second \ixpe epoch, the energy spectra were dominated by the emission from the accretion disk, which, together with the limited fast time variability observed in the power density spectra, indicates that the source had transitioned to the soft state.
    \item We measured a significant polarization in the $3-8$~keV of epoch 1, with $ 1.3\pm0.3\%$  \PDX{} and $-74$\degree$\pm7$\degree \PAX. 
    In the full \ixpe energy band, we only found upper limits of \PDX{} $<1.1\%$ in epoch 1, and $<1.2\%$ in epoch 2 (at 3$\sigma$ confidence level).
    \item We estimated the jet orientation angle on the plane of the sky to be $-69.5$\degree$\pm1.1$\degree, based on the detection of the radio core and a moving discrete jet ejection during our ATCA observation. 
    The \PAX{} measured in epoch 1 is aligned with the direction of the jet.
    \item We analysed three sets of FORS2/VLT observations. We detected significant low  polarization ($<1\%$) in 4 bands (\textit{I} , \textit{R}, \textit{V} and \textit{B}) during all three observations, apart from the \textit{I} band in the last observation which has only an upper limit (see Table~\ref{tab:polla_results}). 
    During the first observation, simultaneous with \ixpe epoch 1, the \PA{} of all 4 bands is compatible with \PAX{} and the jet direction. 
\end{itemize}

Since both the X-ray and optical polarizations are relatively low, we can not exclude any scenario (see Section~\ref{sec:discussion}).  However, based on the  alignment of X-ray and optical \PA{} with the radio jet, we favor a system configuration with the corona horizontally extended on the plane of the accretion disk and the optical radiation produced from the outer regions of the disk. 

\acknowledgments
    We wish to dedicate this paper to our beloved collaborator, mentor to several of us, and friend Tomaso Belloni, who was still with us when this campaign was first conceived.\\ 
    
    GM acknowledges financial support from the European Union’s Horizon Europe research and innovation programme under the Marie Sk\l{}odowska-Curie grant agreement No. 101107057.
    MCB and TDR acknowledge support from the INAF-Astrofit fellowship. FC acknowledges support from the Royal Society through the Newton International Fellowship programme (NIF/R1/211296). FC acknowledge support from the INAF grant "Spin and Geometry in accreting X-ray binaries: The first multi-frequency spectro-polarimetric campaign". YC acknowledges support from the grant RYC2021-032718-I, financed by MCIN/AEI/10.13039/501100011033 and the European Union NextGenerationEU/PRTR.
    BDM acknowledges support via Ram\'on y Cajal Fellowship (RYC2018-025950-I) and the Spanish MINECO grant PID2022-136828NB-C44. BDM and YC thank the Spanish MINECO grant PID2020-117252GB-I00 and the AGAUR/Generalitat de Catalunya grant SGR-386/2021.
    EVL is supported by the Italian Research Center on High Performance Computing Big Data and Quantum Computing (ICSC), project funded by European Union - NextGenerationEU - and National Recovery and Resilience Plan (NRRP) - Mission 4 Component 2 within the activities of Spoke 3 (Astrophysics and Cosmos Observations). SF acknowledges the support by the project PRIN 2022 - 2022LWPEXW - “An X-ray view of compact objects in polarized light”, CUP C53D23001180006. DMR, PS and KA are supported by Tamkeen under the NYU Abu Dhabi Research Institute grant CASS. MD and JS thank GACR project 21-06825X for the support and institutional support from RVO:67985815.
    AM is supported by the European Research Council (ERC) under the European Union’s Horizon 2020 research and innovation programme (ERC Consolidator Grant "MAGNESIA" No. 817661, PI: Rea).   
    PC acknowledges financial support from the Italian Space Agency and National Institute for Astrophysics, ASI/INAF, under agreement ASI-INAF n.2017-14-H.0
    ATCA is part of the Australia Telescope National Facility (https://ror.org/05qajvd42) which is funded by the Australian Government for operation as a National Facility managed by CSIRO. We acknowledge the Gomeroi people as the Traditional Owners of the ATCA observatory site. 
    Part of the observations presented in this study were obtained at the European Southern Observatory under ESO programme 112.25UU.001 (PI: Baglio).
    The Authors thank the Team Meetings at the International Space Science Institute (Bern) for fruitful discussions and were supported by the ISSI International Team project \#440.



\bibliography{library2024}{}
\bibliographystyle{aasjournal}

\appendix
\section{The outburst of \source}
\label{app:gx_outbursts}

\begin{figure} 
\centering
\includegraphics[width=\linewidth]{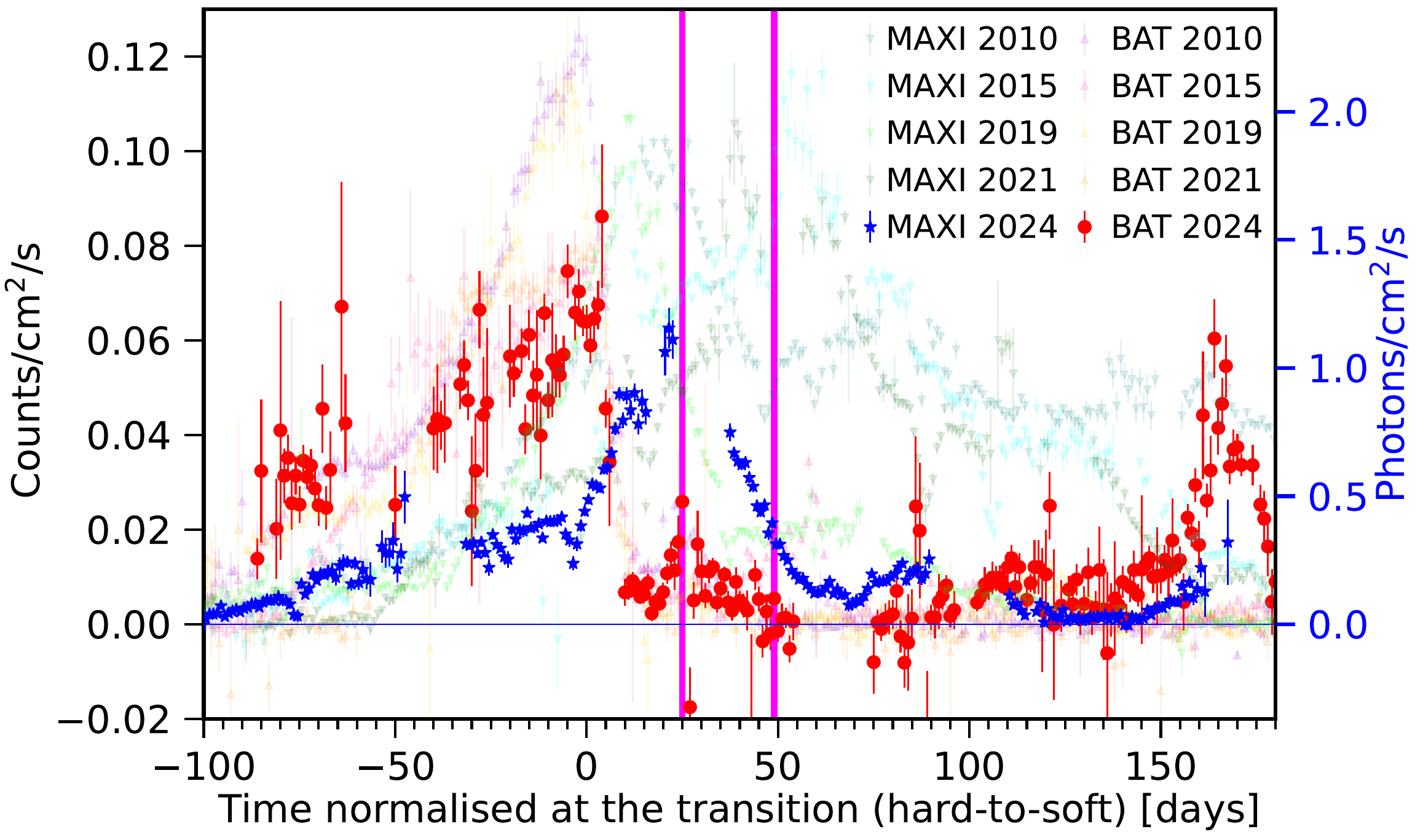}
\caption{\maxi\ and Swift/BAT light curves of \source\ outbursts of 2010, 2015, 2019, 2019 and 2024. All the outbursts are synchronized at the beginning of the state transition (day 0). The vertical magenta lines indicate the two epochs discussed in this work for the 2024 outburst (Feb 14, 2024, and Mar 8, 2024).} 
\label{fig:MAXI_BAT_lc}
\end{figure}

\source\ was in outburst 5 times in the last 15 years (not counting the hard-only outbursts). 
Figure~\ref{fig:MAXI_BAT_lc} shows the \maxi\ and Swift/BAT light curves highlighting the 2024 outburst that we consider in our work. 
\maxi\ traces the soft flux since its energy range is 2-20~keV, and Swift/BAT observes the hard flux (15-150~keV).
All the light curves are aligned at the beginning of the hard-to-soft transition, when the hard flux starts to drop in Swift/BAT. 
The 2024 outburst presents one of the lowest fluxes among the last outbursts, especially in the soft state.  

\begin{figure} 
\centering
\includegraphics[width=\linewidth]{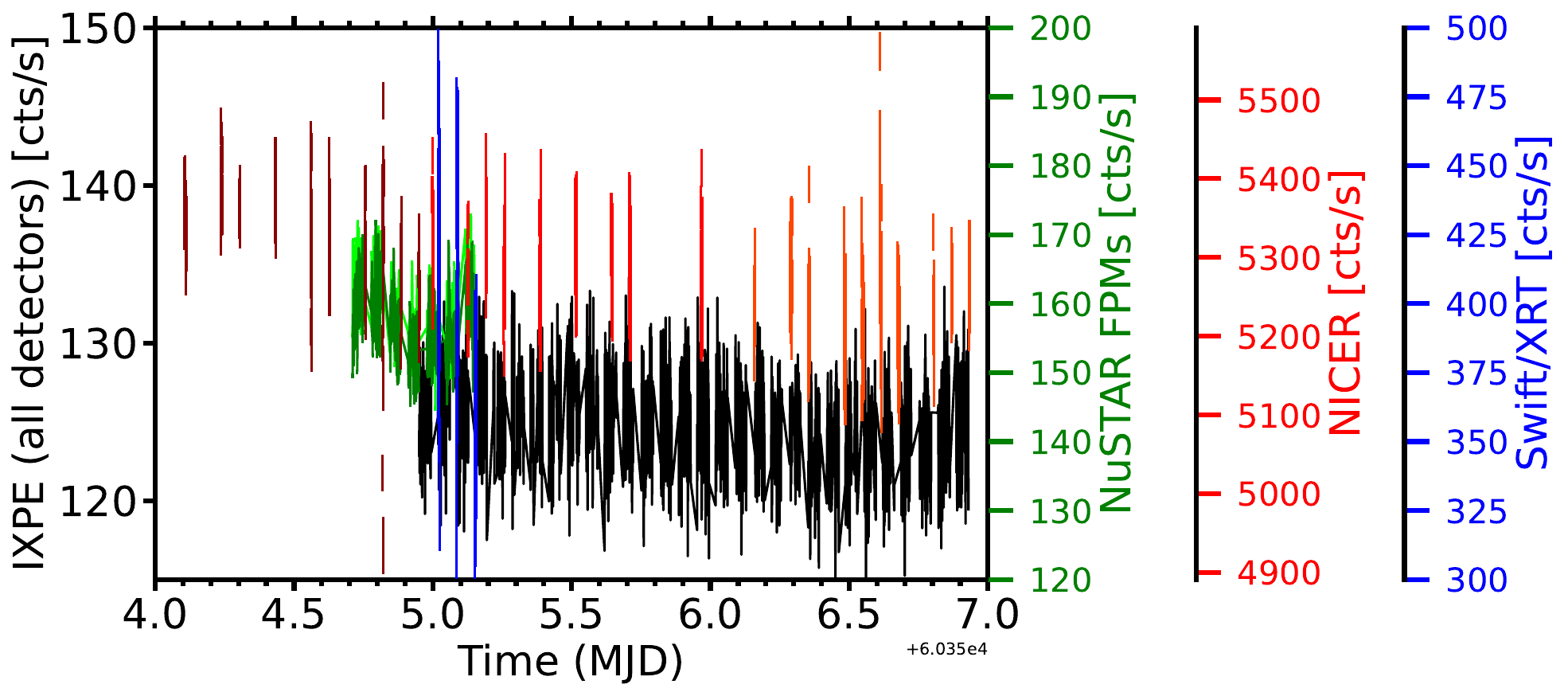}
\includegraphics[width=\linewidth]{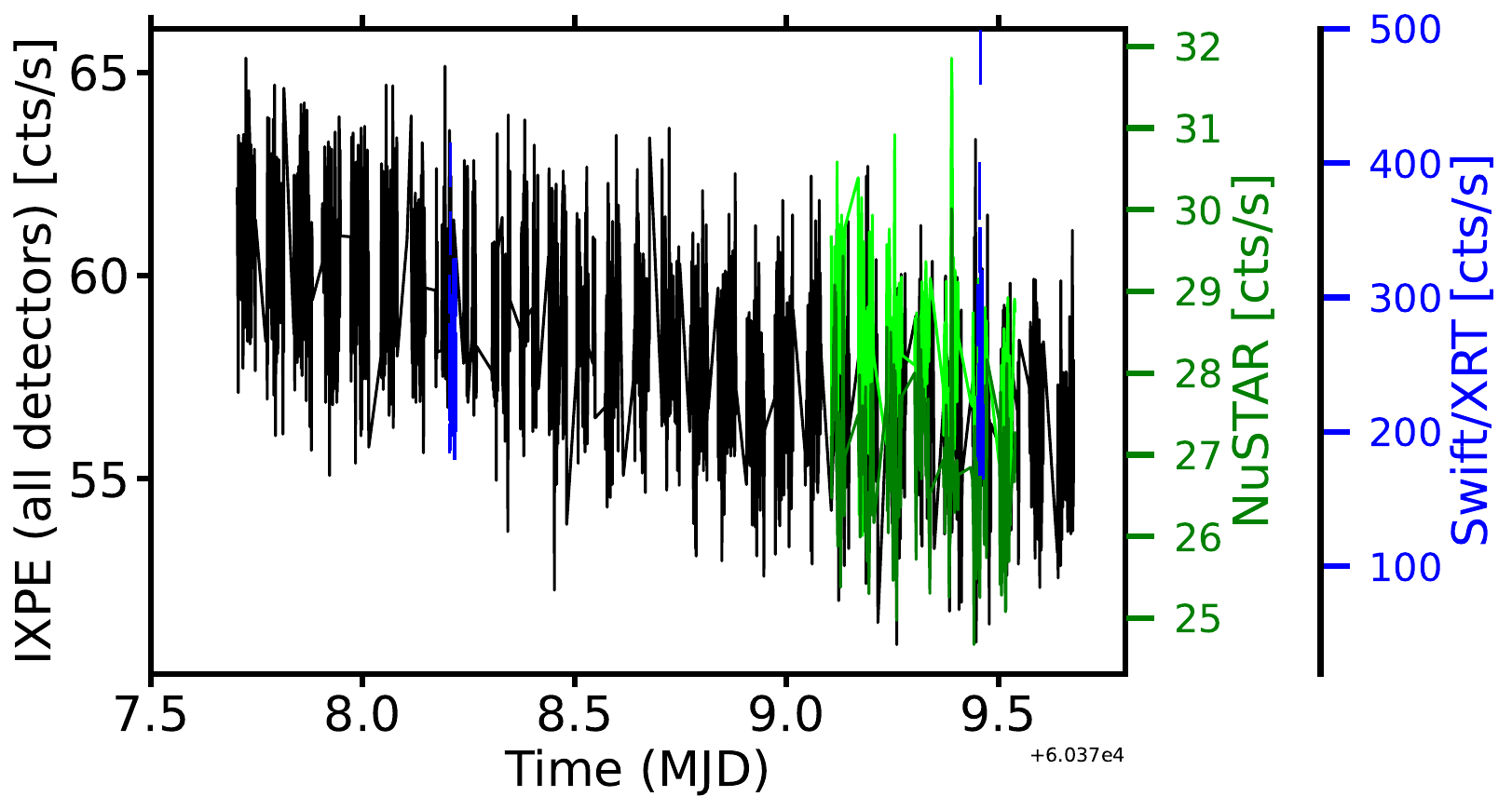}
\caption{Light curves of all the X-ray instruments considered in this work during epoch 1 (top panel) and epoch 2 (bottom panel). Each instrument has its own y-axis showing the count rate. \ixpe (black points) is on the left, while the indicators of the count rate for all the other instruments are on the right (\nustar in green, \swiftxrt in blue and \nicer in red).} \label{fig:lc}    
\end{figure}

During the two \ixpe epochs \source\ was observed by a few other X-ray telescopes. 
Figure~\ref{fig:lc} shows the light curves either simultaneous or partially simultaneous to the two \ixpe observations. \ixpe and \nustar lightcurves have 50 seconds time resolution, \nicer and \swiftxrt lightcurves have 10 seconds time resolution. 
We note that we combined the counts of all three \ixpe detectors. 
During epoch 1 the source flux does not show strong long timescale variability. 
\nustar and \swiftxrt observed just in the first portion of the first \ixpe epoch, whereas \nicer overlaps with almost the whole \ixpe exposure. 
During the second epoch, we note that the final flux has decreased by roughly 10\% of its initial value. 
\nustar observed at the end of the \ixpe observation together with one of the two \swiftxrt observations. 

\subsection{All the observations}
\ixpe observed \source\ twice during 2024. 
\swiftxrt, \nustar and \nicer observed the source simultaneously to the \ixpe epochs.  
The exact dates of the observations that we considered are reported in Table~\ref{tab:all_obs}.
The \nicer Guest Observer program 7702010xxx was also used to daily monitor the source from January 31, 2024, to February 21, 2024. 
Table~\ref{tab:all_obs} shows all the \nicer observations of our campaign. 
When an observation is reported with 0~s exposure, it means that it was performed in orbit day. 
Due to the optical light leak that is affecting the \nicer instruments, these observations cannot be used for scientific analysis. 
The monitoring of the source started on January 31, 2024, with an almost daily cadence until February 21, 2024. 
In particular, obsID 7702010111, 7702010112, and 7702010113 are the observations simultaneous to \ixpe epoch 1.
The monitoring then resumed on March 8, 2024. These last three observations are simultaneous to \ixpe epoch 2, but they turned out to be heavily corrupted by the light leak. 

\begin{table*}
\centering
\setlength{\tabcolsep}{8pt} 
\renewcommand{\arraystretch}{1.5} 
\begin{tabular}{ c | c | c | c | c}

\hline
Telescope & ObsID & Start & End & Exposure (s) \\
\hline
\ixpe& 03005101 & 2024 Feb 14\quad 22:48:57 & 2024 Feb 16\quad 22:19:34  & 94600 \\
    & 03005301 & 2024 Mar 08\quad 16:53:26 & 2024 Mar 10\quad 16:16:18  & 97700 \\
\hline
\swiftxrt & 00014052207 & 2024 Feb 15\quad 00:25:11 & 2024 Feb 15\quad 00:31:56 & 404\\
 & 00014052208 & 2024 Feb 15\quad 02:00:20 & 2024 Feb 15\quad 02:06:55& 394 \\
 & 00014052209 & 2024 Feb 15\quad 03:35:11 & 2024 Feb 15\quad 03:41:55 & 404 \\
 & 00014052219 & 2024 Mar 09\quad 04:51:32 & 2024 Mar 09\quad 05:14:56 & 1400 \\
 & 00016552003 & 2024 Mar 10\quad 10:47:44 & 2024 Mar 10\quad 11:04:56 & 1028 \\
 
\hline
\nustar & 91002306002 & 2024 Feb 14\quad 16:51:09 & 2024 Feb 15\quad 04:01:09 & 16400\\
       & 80902342002 & 2024 Mar 10\quad 02:31:09 & 2024 Mar 10\quad 13:21:09 & 19500\\
\hline
\nicer & 7702010101 & 2024 Jan 31\quad 16:04:28 & 2024 Jan 31\quad 16:13:13 & 525 \\
& 7702010102 & 2024 Feb 01\quad 07:34:25 & 2024 Feb 01\quad 10:48:17 & 1438 \\
& 7702010104 & 2024 Feb 07\quad 13:44:26 & 2024 Feb 07\quad 13:52:40 & 490 \\
& 7702010105 & 2024 Feb 08\quad 05:16:24 & 2024 Feb 08\quad 22:30:11 & 3730 \\
& 7702010106 & 2024 Feb 08\quad 23:55:03 & 2024 Feb 09\quad 18:40:40 & 1679 \\
& 7702010107 & 2024 Feb 10\quad 07:01:06 & 2024 Feb 10\quad 14:51:53 & 636 \\
& 7702010108 & 2024 Feb 11\quad 01:41:39 & 2024 Feb 11\quad 20:22:13 & 954 \\
& 7702010109 & 2024 Feb 12\quad 04:04:33 & 2024 Feb 12\quad 04:07:58 & 205 \\
& 7702010110 & 2024 Feb 13\quad 14:04:21 & 2024 Feb 13\quad 20:28:19 & 986 \\
(Epoch 1) & 7702010111 & 2024 Feb 14\quad 02:30:27 & 2024 Feb 14\quad 22:49:57 & 3465 \\
(Epoch 1) & 7702010112 & 2024 Feb 14\quad 23:55:27 & 2024 Feb 15\quad 23:17:40 & 3802 \\
(Epoch 1) & 7702010113 & 2024 Feb 16\quad 03:46:45 & 2024 Feb 16\quad 22:26:00 & 4778 \\
& 7702010114 & 2024 Feb 17\quad 12:17:23 & 2024 Feb 17\quad 13:56:00 & 601 \\
& 7702010115 & 2024 Feb 18\quad 13:03:25 & 2024 Feb 18\quad 13:12:23 & 538 \\
& 7702010116 & 2024 Feb 19\quad 15:25:03 & 2024 Feb 19\quad 15:35:40 & 635 \\
& 7702010117 & 2024 Feb 20\quad 14:43:29 & 2024 Feb 20\quad 16:22:00 & 660 \\
& 7702010118 & 2024 Feb 21\quad 00:03:53 & 2024 Feb 21\quad 01:39:20 & 261 \\
(Epoch 2) & 7702010119 & 2024 Mar 08\quad 13:02:44 & 2024 Mar 08\quad 13:02:44 & 0 \\
(Epoch 2) & 7702010120 & 2024 Mar 08\quad 23:54:45 & 2024 Mar 08\quad 23:54:45 & 0 \\
(Epoch 2) & 7702010121 & 2024 Mar 10\quad 00:40:51 & 2024 Mar 10\quad 00:40:51 & 0 \\
\hline

\end{tabular}
\caption{The dates of the observations by \ixpe, \swiftxrt, \nustar and \nicer used in this paper.}
\label{tab:all_obs}
\end{table*}

\subsection{FORS2 analysis}
\label{app:optical_pol_data}
VLT/FORS2 observed \source\ three nights during the 2023/2024 outburst, February 15, March 2, and March 16, 2024, as part of a larger optical polarization campaign on the source (Baglio et al. in preparation). 
We consider the first observation which is simultaneous to \ixpe epoch 1 and the last two observations which are the closest to \ixpe epoch 2. 
Aperture photometry was conducted with the DAOPHOT tool \citep{Stetson1987}, using a 6-pixel aperture. 
The normalized Stokes parameters Q and U for linear polarization were calculated following the methods outlined by \cite{Baglio2020} (eq. 1 - 2). 
We note that these parameters are not corrected for instrumental contributions to the linear polarization. 
However, unpolarized standard stars are regularly observed with FORS2 to monitor the level of instrumental polarization, which has remained stable over the past 10 years across all bands at a very low level ($<0.3\%$). 
We then used the algorithm described in \cite{Baglio2020} and references therein to evaluate the linear polarization of \source\ starting from the parameter $S(\Phi)$ for each HWP angle, defined as:

\begin{equation}
S(\Phi)=\left( \frac{f^{o}(\Phi)/f^e(\Phi)}{f^o_u(\Phi)/f^e_u(\Phi)}-1\right)/\left( \frac{f^{o}(\Phi)/f^e(\Phi)}{f^o_u(\Phi)/f^e_u(\Phi)}+1\right),
\end{equation}
where $f^o(\Phi)$ and $f^e(\Phi)$ are the ordinary and extraordinary fluxes of \source, respectively, and $f^o_u(\Phi)$ and $f^e_u(\Phi)$ are the same quantities calculated for an unpolarized standard star in the field.
This parameter is linked to the polarization \PD{} of the target and to its polarization angle \PA{} by the following equation:
\begin{equation}\label{eq_cos}
S(\Phi) = \PDe\, \cos 2(\PAe - \Phi).
\end{equation}

Therefore, a fit of the $S$ parameter with Equation~\ref{eq_cos} will give an estimate of the linear \PD{} and \PA{} for the target.
To increase the significance of the fit, we considered 10 reference field stars in each epoch. 
Under the simple hypothesis that the field stars are intrinsically unpolarized, this method gives as a result a linear polarization for the target that is already corrected for the low instrumental effects. 
In addition, if the Q and U Stokes parameters of the reference stars are consistent with each other, this method should, in principle, automatically correct for interstellar polarization along the line of sight. 
Unfortunately, \source\ is in a highly absorbed region \citep[$A_V=3.58$ mag, where $A_V$ is the absorption coefficient in $V$-band;][]{Kosenkov2020}, and the distance of the source is likely high \citep[$10\pm2$ kpc;][]{Zdziarski2019}; therefore, it is possible that, despite the correction, some residual interstellar polarization is still present in our results. We note, however, that the interstellar polarization in the direction of \source\ has been estimated in the past by \cite{Russell2008} thanks to polarimetric observations of the source performed close to quiescence. The polarization angle of interstellar dust polarization was estimated to be $\sim 30^{\circ}$: this is not consistent with the (interstellar dust-subtracted) polarization angle measured in this work, which is instead parallel to the jet axis of \source. It is therefore unlikely that the optical linear polarization reported in this work has a dominant interstellar origin.

Following \citet{Baglio2020}, to evaluate \PD{} and \PA{} we maximized the Gaussian likelihood function using an optimization algorithm \citep[e.g., the Nelder–Mead algorithm;][]{Gao&Han2012} and integrated the posterior probability density of our model parameters using a Markov Chain Monte Carlo (MCMC) algorithm \citep{Hogg&Foreman2018} based on the ``affine-invariant Hamiltonia'' algorithm \citep{Foreman-Mackeyetal2013}. The chains were initiated from small Gaussian distributions centered on the best-fit values. 
We discarded the first third of each chain as the ``burn-in phase'' and ensured that a stationary distribution was reached \citep{Sharma2017}. 
The quality of the fit was assessed as described in \cite{Lucy2016}. 
The values for \PD{} and \PA, along with their $1\sigma$ uncertainties, correspond to the 0.16, 0.50, and 0.84 quantiles of the posterior distribution of the parameters. In the case of non-detections, the 99.97\% percentile of the posterior distribution of the parameter \PD{} was used to estimate an upper limit.
The value of \PA{} derived using this method was further adjusted based on observations of the polarized standard star Vela 1-95, with known and documented polarization angle in all FORS2 bands. 
The average correction applied was negligible, remaining under $2^{\circ}$ across all bands and epochs.

\section{Spectro-polarimetric fit in details}
\label{app:spectral_fit}
We simultaneously fit the energy spectra of the three \swiftxrt observations and the \nustar observation in epoch 1 with the second \swiftxrt observation and the \nustar observation in epoch 2. 
We did not use the \nicer spectra since we could use them only for epoch 1 (see Section~\ref{sec:Obs}).
However, we performed some preliminary analysis fitting combinations of the \nicer and \nustar data,  Swft/XRT data and \nustar data: all these preliminary fits of epoch 1 were extremely similar.
Our best fit model is \texttt{TBnew\_feo(thcomp*kerrbb + relline)*smedge}. 
All the absorption parameters, the spin of the black hole and the inclination of the system are tied between the two epochs. 
Regarding \texttt{kerrbb}, we choose a standard Keplerian disk with zero torque at the inner boundary (eta = 0.0). 
We also fix the spectral hardening (hd) to 1.7 \citep{Shimura1995}, the distance to 10~kpc and the black hole mass to 10~$M_\odot$. 
We include limb-darkening and we do not include self-irradiation in the disk calculation of the \texttt{kerrbb} (rflag = 1 and lflag = 0). 
All the other parameters are free to vary, however, only the mass accretion rate and the normalization vary between the two epochs in \texttt{kerrbb}.
We use the \texttt{thcomp} convolution model to fit the Comptonization component in the spectrum, all its parameters are free to vary between the two epochs since the corona is supposed to change in different states. 
The only exception is the Cosmological redshit which is set to zero. 
The \texttt{relline} model is used to fit the iron K$\alpha$ line. 
Since most of the advanced reflection models (such as the \texttt{relxill} suit of models) do not include self-irradiation from the accretion disk, which should be relevant when the source is in the soft-intermediate state and in the soft state, we decided not to use them. 

\begin{figure} 
\centering
\includegraphics[width=\linewidth]{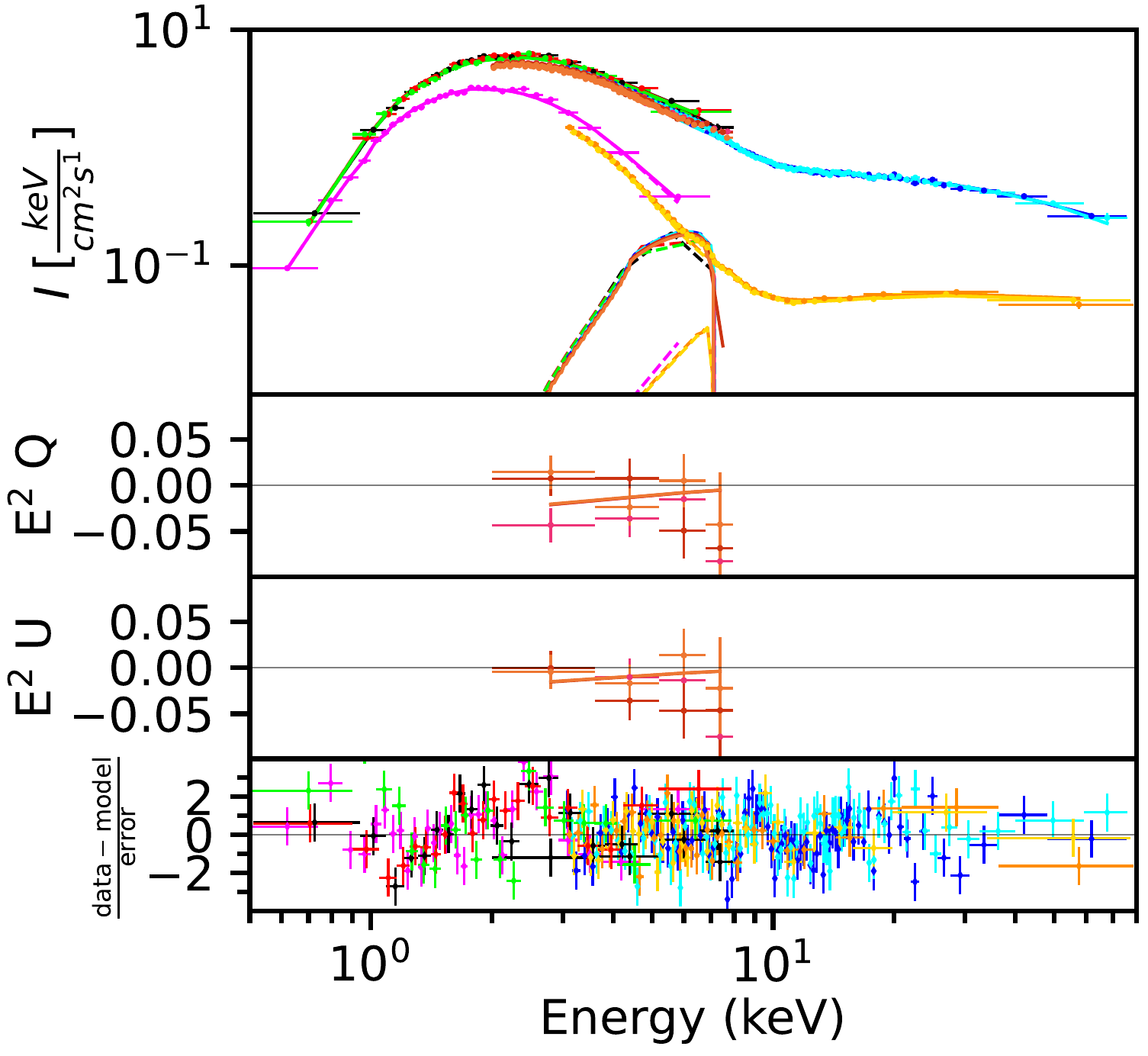}
\caption{Top: unfolded spectra of the two epochs. Spectra of epoch 1:  \swiftxrt (black, green, and red) and \nustar FPMA and FPMB (blue and cyan). Spectra of epoch 2: \swiftxrt (magenta) and \nustar FPMA and FPMB (yellow and orange). Central panels: energy spectrum of Stokes parameters Q and U of epoch 1 only. Bottom: spectral fit residuals of only \swiftxrt and \nustar spectra for both epochs.} \label{fig:spectro_pol_fit}
\end{figure}

We fit the \swiftxrt spectra with a gain shift (\texttt{xspec} command \textit{gain fit}). 
We allow both slope and offset to vary during the fit, however, we tie these two parameters between the three \swiftxrt observations of epoch 1. 
Using a gain shift is suggested by the International Astronomical Consortium for High-Energy Calibration (IACHEC).
In particular, \citet{Madsen2017crosscal} reported that \swiftxrt high signal-to-noise spectra of bright sources occasionally show residuals of 10\% level, suggesting the use of the \textit{gain} command in \texttt{xspec} to mitigate this effects. 
We note that the fit requires higher values of the gain offset, $84\pm20$~eV for epoch 1 and $140\pm20$~eV\footnote{Errors are quoted at $3\sigma$.} compared to what is suggested by \citet{Madsen2017crosscal} ($\sim\pm10-50$~eV).
Table~\ref{tab:bestfit_par} shows the best-fit parameter values. 
The reduced $\chi^2$ is 2461 over 2254 degrees of freedom (d.o.f.). 
Figure~\ref{fig:spectro_pol_fit} shows the unfolded spectra of the two epochs:  black, green, and red symbols are the three \swiftxrt energy spectra of epoch 1, blue and cyan are the FPMA and FPMB \nustar spectra of epoch 1; magenta is the \swiftxrt spectrum of epoch 2, yellow and orange are the FPMA and FPMB \nustar spectra of  epoch 2. 
The residuals match the colors of the spectra. 
We note that the residuals show a few features around low energies (below 3~keV), this might be due to possible instrumental features which have been reported by the calibration team\footnote{www.heasarc.gsfc.nasa.gov/docs/heasarc/caldb/swift/docs/xrt/SWIFT-XRT-CALDB-09\_v19.pdf}  due to the high-energy proton interactions causing damage to the CCD\footnote{For example, near the Au–M V edge at 2.205~keV, the Si–K edge at 1.839~keV, or the O–K edge at 0.545~keV \citep[see][]{Madsen2017crosscal}.}. 

\begin{table}
\centering
\setlength{\tabcolsep}{8pt} 
\renewcommand{\arraystretch}{1.5} 
\caption{Best fit parameters of the joint fit between the first and the second epochs. The model is \texttt{constant*TBnew\_feo*(thcomp*kerrbb + relline)*smedge}, we fix eta = 0.0, hd = 1.7 in \texttt{kerrbb}, Index 2 = 3.0, Rbr = 15~R$_{\rm g}$, inner radius = ISCO, the outer radius = 400~R$_{\rm g}$ in  \texttt{relline}, and index = -2.67 in \texttt{smedge}. The errors are at the 90\% confidence level. }

\begin{tabular}{ c c c c }
\hline
\hline

\multicolumn{4}{c}{Spectral fit: \swiftxrt and \nustar}\\

\hline
 &  & \multicolumn{2}{c}{Epoch 1 \& 2} \\
\hline
\hline
tbnew\_feo & nH [$10^{22}\, {\rm cm^{-2}}$]  & \multicolumn{2}{c}{$       0.4^{+0.1}_{-0.1}$} \\
           &  O   & \multicolumn{2}{c}{$       2.2^{+0.8}_{-0.8}$  } \\
           & Fe   & \multicolumn{2}{c}{$       1.3^{+0.4}_{-0.4}$  } \\
\hline
kerrbb     & a    & \multicolumn{2}{c}{$       0.85^{+0.01}_{-0.01}$} \\
           & incl [ \degree] & \multicolumn{2}{c}{$      30^{+2}_{a}$         } \\
\hline
\hline
           &           & Epoch 1                          & Epoch 2 \\
\hline
\hline
kerrbb  &  Mdd [$10^{18} g/s$] & $       1.06^{+0.13}_{-0.04}$      & $       0.6^{+0.5}_{-0.2}$ \\
        &  norm        & $       1.7^{+0.2}_{-0.1}$       & $       1.9^{+0.2}_{-0.2}$ \\
\hline
thcomp  & $\Gamma$     & $       2.35^{ +0.05}_{-0.02}$   & $       2.12^{+0.03}_{-0.03}$    \\
        & kT$_{\rm e}$ [keV] & $      30^{+6}_{-8}$             & $      797^{b}$             \\
        & cov\_frac    & $       0.17^{+0.02} _{-0.01}$   & $       0.029^{+0.002}_{-0.014}$ \\
\hline
relline & lineE [keV]  & $       6.84^{+0.03}_{-0.06}$       & $       6.71^{+0.06}_{-0.07}$       \\
        & Index1       & $       4.48^{+0.04}_{-0.13}$    & $       3.3^{+0.3}_{-0.6}$       \\
        & norm         & $       0.022^{+0.001}_{-0.003}$ & $       0.002^{+0.001}_{-0.001}$ \\
\hline
smedge  & edgeE [keV]  & $       7.76^{+0.04}_{-0.07}$    & $       7.8^{+0.1}_{-0.1}$ \\
        & MaxTau       & $       8^{+1}_{-3} $                   & $  2.8^{c}$             \\
        & width [keV]  & $      39^{+6}_{-10}$   & $      8^{+6}_{-3}$     \\
\hline
Swift Obs2 & cal const & $       1.00^{+0.01}_{-0.01}$    & -- \\
\hline
Swift Obs3 & cal const & $       0.98^{+0.01}_{-0.01}$    & -- \\
\hline
gain fit -- Swift & slope & $1.06^{+0.01}_{-0.01}$ & $1.058^{+0.010}_{-0.004}$ \\
& offset [eV] & $-84^{+11}_{-10}$ & $-139^{+12}_{-12}$\\

\hline
\nustar FPMA   & cal const               & $ 0.91^{+0.01}_{-0.01}$ & $ 0.75^{+0.02}_{-0.01}$ \\
\hline
\nustar FPMB   & cal const               & $ 0.90^{+0.01}_{-0.02}$ & $ 0.73^{+0.02}_{-0.01}$ \\
\hline
           & $\chi^2 / {d.o.f.}$ & \multicolumn{2}{c}{ 2461 / 2254 } \\

\hline
\hline

\end{tabular}
	\begin{list}{}{}
    	\item[$^*$] The systematics are not accounted for by the Cash statistic, thus the statistical errors quoted are unrealistically small. 
    	\item[$^a$] The lower limit of the inclination is 30. 
    	\item[$^b$] The electron temperature of epoch 2 is not constrained (fit range $0.5-900$~keV).
 		\item[$^c$] The MaxTau value is not constrained (fit range $0-5$).
	\end{list}
\label{tab:bestfit_par}
\end{table}

After we establish the best-fit, we add the \ixpe spectra \textit{I}, \textit{Q} and \textit{U} for each of the three \ixpe detectors allowing us to measure the polarization of the source. 
We freeze all the physical parameters to the best-fit values and leave the calibration constants free to vary among all the instruments. 
We note that the \texttt{gain fit} functionality of \texttt{xspec} is required to fit the energy spectra of the three \ixpe detectors.  
Figure~\ref{fig:spectro_pol_fit} shows the three \ixpe \textit{I} spectra of epoch 1 in the top panel (dark red, light red, and orange colors for detector 1, 2 and 3 respectively) and the \textit{Q} and \textit{U} spectra of each detectors in the middle panels (matching the colors of the \textit{I} spectra). 
We apply the \texttt{polconst} multiplicative model to the entire spectral model. 
As we explained in the main text, we measure significant \PD{} only when we consider the $3-8$~keV band of the \ixpe Stoke parameter energy spectra. 
We analyzed \ixpe epoch 2 with the exact same procedure. 
In Table~\ref{tab:bestfit_pol_constant} we report the polarization best-fit values of \PDX{} and \PAX{}, along with the values of the calibration constants, the gain curve and the pointing offset corrections in the $3-8$~keV energy band for epoch 1 and in the full \ixpe energy band ($2-8$~keV) for epoch 2. 

In order to allow the spectral components to have different polarization values we cannot use the convolution model \texttt{thcomp}. 
Therefore, even though we lose self consistency, we fit the data with the model \texttt{TBnew\_feo*(diskbb + nthcomp + relline)*smedge}. 
We follow the same procedure as in the previous case, finding the best-fit without \ixpe spectra, and freezing the values of the parameters apart from the calibration constants among the instruments. 
In this configuration, we multiply each of the additive components of the model by \texttt{polconst} and fix the \PDX{} of the iron line to zero. 
The line is not supposed to be polarized \citep{Podgorny2022}. 
We first fit allowing either the disk component or the Comptonization component to be polarized. 
In the first case the polarization of the disk component is \PDXdisk{} = $2.0\%\pm0.5\%$ and \PAXdisk{} = -75\degree$\pm $7\degree, while in the second case the polarization of the Comptonization component is \PDXnth{} = $3.6\%\pm0.1\%$ and \PAXnth{} = -73\degree$\pm $7\degree (1$\sigma$ errors).

\begin{figure} 
\centering
\includegraphics[width=\linewidth]{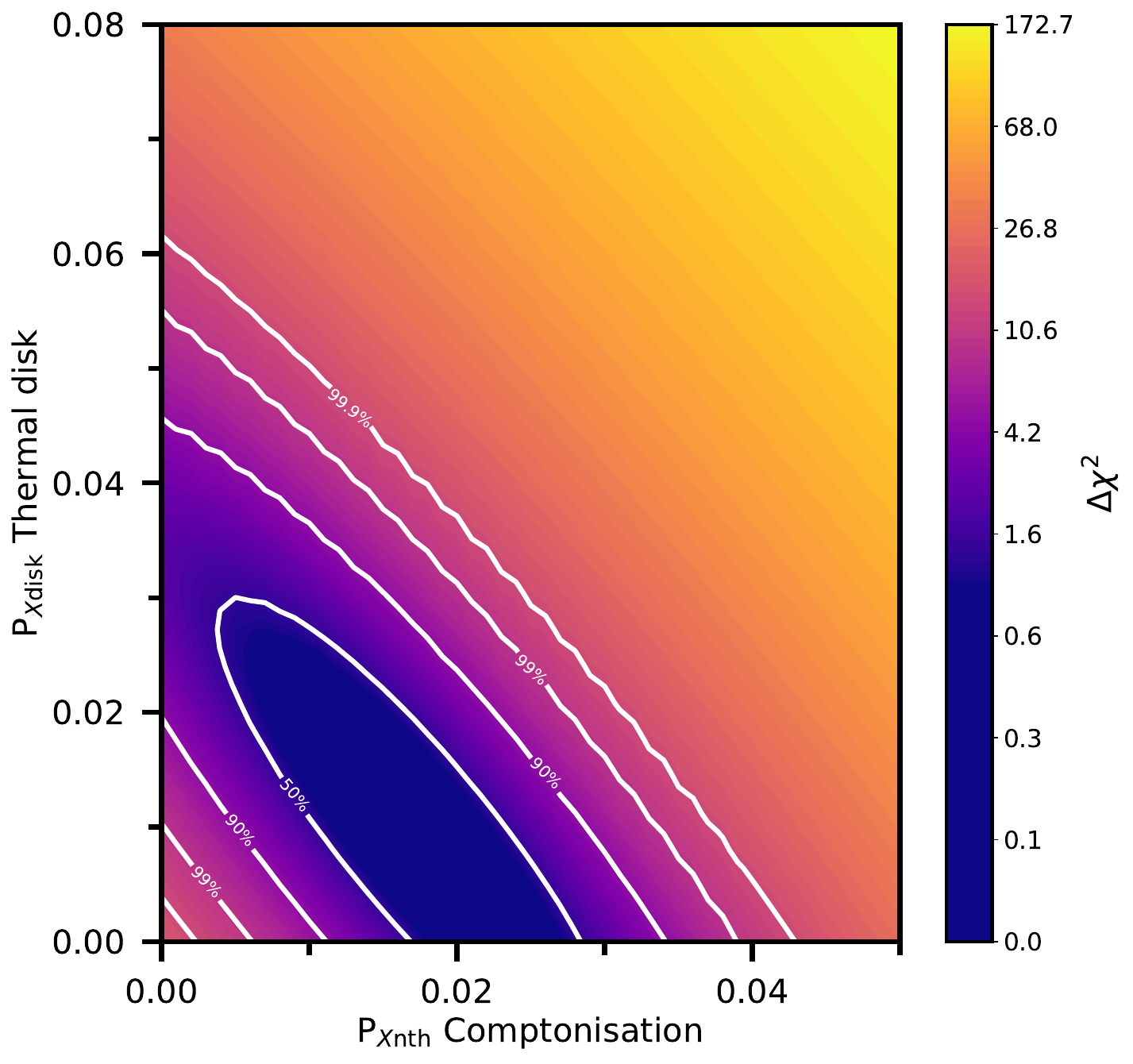}
\caption{Epoch 1 contour plot of the disk component polarization fraction \PDXdisk vs the Comptonization component polarization fraction \PDXnth, when the \PAX{}s of the two components are forced to be aligned.} \label{fig:Xpol_components}    
\end{figure}

\begin{figure} 
\centering
\includegraphics[width=\linewidth]{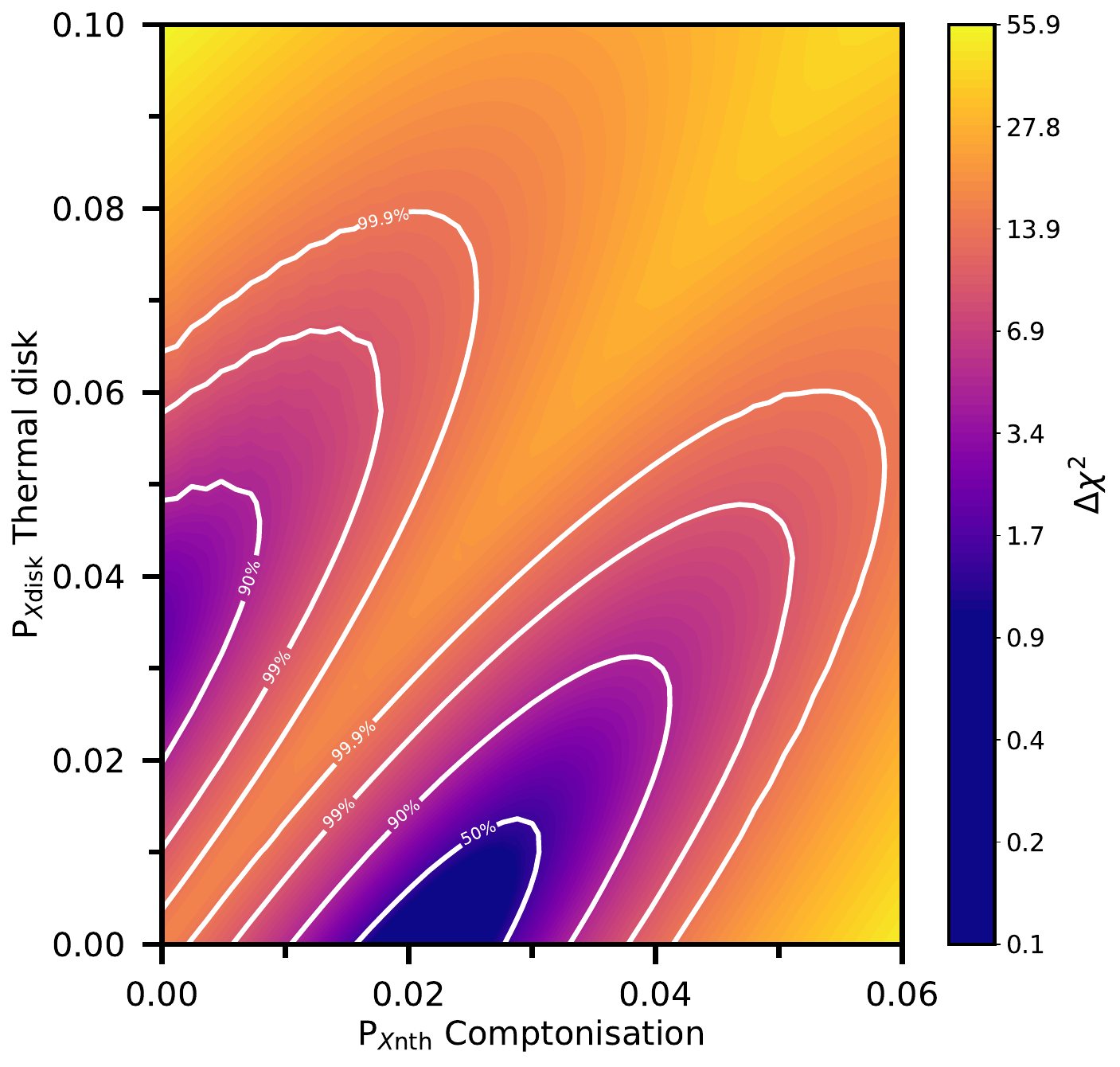}
\caption{Epoch 1 contour plot of the disk component polarization fraction \PDXdisk vs the Comptonization component polarization fraction \PDXnth, when the \PAX{}s of the two components are forced to be perpendicular to each other.} \label{fig:Xpol_components2}    
\end{figure}

Finally we test two scenarios: first we allow \PDX{} of both components to be free and we tie the \PAXdisk{} and \PAXnth{} to be the same; second we force the \PAXdisk{} to be 90\degree off from the \PAXnth{} value which is free to vary in the fit. 
Figure~\ref{fig:Xpol_components} and Figure~\ref{fig:Xpol_components2} show the contour plots of the two \PDX{} components in the first and the second test, respectively. 
 It is interesting to note that, when the polarization angles are forced to be perpendicular, only one component can have a strong polarization degree, further strengthening the conclusion that the polarization angle can only be $\sim -74$\degree.
\begin{table}
\centering
\setlength{\tabcolsep}{8pt} 
\renewcommand{\arraystretch}{1.5} 
\caption{Spectro-polarimetric calibration constants. The errors are at the 90\% confidence level apart from \PDX{} and \PAX{} that are spcified in the table. }

\begin{tabular}{ c c c c }
\hline
\hline

\multicolumn{4}{c}{Calibration parameters for Spectro-polarimetric fit$^{f}$}\\

\hline
\hline
           &           & Epoch 1    [$3-8$~keV]      & Epoch 2 [$2-8$~keV]\\
\hline
& \PDX{} &$ 1.3\pm0.3\%$ (1$\sigma$)   &  $<1.2\%$ (3$\sigma$) \\
&\PAX{} & $-74$\degree$\pm7$\degree (1$\sigma$) & -- \\
\hline
\ixpe det1 & cal const & $0.876^{+0.004}_{-0.004}$ & $0.793^{+0.007}_{-0.006}$\\

gain fit  & slope & $0.967^{+0.002}_{-0.002}$ & $0.984^{+0.003}_{-0.003}$ \\
& offset [eV] & $30^{+7}_{-7}$ & $7^{+12}_{-11}$ \\
\hline
\ixpe det2 & cal const & $0.873^{+0.004}_{-0.004}$ & $0.794^{+0.007}_{-0.007}$\\
gain fit   & slope & $0.975^{+0.002}_{-0.002}$ & $0.965^{+0.003}_{-0.003}$ \\
& offset [eV] & $28^{+8}_{-7}$ & $52^{+12}_{-13}$ \\
\hline
\ixpe det3 & cal const & $0.848^{+0.004}_{-0.004}$ & $0.760^{+0.007}_{-0.007}$ \\
gain fit   & slope & $0.972^{+0.002}_{-0.002}$ & $0.982^{+0.003}_{-0.003}$ \\
& offset [eV] & $28^{+7}_{-8}$ & $23^{+12}_{-12}$ \\
\hline
           & $\chi^2 / {d.o.f.}$ & 3871 / 3611 & 3799 / 3611 \\
\hline

\end{tabular}
	\begin{list}{}{}
    	\item[$^f$] All the previous parameters are fixed apart from calibration constants and gain fit parameters. 
	\end{list}
\label{tab:bestfit_pol_constant}
\end{table}

\end{document}